\documentclass[final,5p,twocolumn,10pt]{elsarticle}

\usepackage[subpreambles=true]{standalone}
\usepackage{import}

\usepackage{amssymb}
\usepackage{amsmath}
\usepackage{color}
\usepackage{physics}
\usepackage{booktabs}

\usepackage{multirow}
\usepackage{array}

\usepackage[hidelinks]{hyperref}

\newcommand{\microg}{$\mu$g}
\newcommand{\sub}[1]{_{\text{#1}}}

\newcommand{\rev}[1]{\textcolor{black}{#1}}

\journal{Computers in Biology and Medicine}

\newenvironment{creditlist}{
	\newcommand{\credit}[2]{\textbf{##1:} ##2.}
}

\begin{document}

\begin{frontmatter}



\title{In silico evaluation of pramlintide dosing algorithms in artificial pancreas systems} 


\author[i2mb]{Borja Pons Torres}
\ead{borponto@etsii.upv.es}
\author[ai2,ciberdem]{Iván Sala-Mira}
\ead{ivsami@upv.es}
\author[ai2]{Clara Furió-Novejarque}
\ead{clafuno@upv.es}
\author[uv]{Ricardo Sanz}
\ead{ricardo.sanz@uv.es}
\author[ai2,ciberdem]{Pedro García}
\ead{pggil@isa.upv.es}
\author[ai2,ciberdem]{José-Luis Díez}
\ead{jldiez@isa.upv.es}
\author[ai2,ciberdem]{Jorge Bondia\corref{cor1}} 
\ead{jbondia@isa.upv.es}

\cortext[cor1]{Corresponding author}

\affiliation[i2mb]{organization={Instituto Universitario de Investigación Concertado de Ingeniería Mecánica y Biomecánica, Universitat Politècnica de València},
	city={València},
	country={Spain}}
	
\affiliation[ai2]{organization={Instituto Universitario de Automática e Informática Industrial, Universitat Politècnica de València},
            city={València},
            country={Spain}}
            
\affiliation[ciberdem]{organization={Centro de Investigación Biomédica en Red de Diabetes y Enfermedades Metabólicas Asociadas, Instituto de Salud Carlos~III},
	city={Madrid},
	country={Spain}}

\affiliation[uv]{organization={Department of Electronic Engineering, University of Valencia},
	city={Burjassot},
	country={Spain}}

\begin{abstract}
Pramlintide's capability to delay gastric emptying has motivated its use in \rev{artificial pancreas systems}, accompanying insulin as a control action. Due to the scarcity of pramlintide simulation models in the literature, in silico testing of insulin-plus-pramlintide strategies is not widely used. This work incorporates a recent pramlintide pharmacokinetics/pharmacodynamics model into the T1DM UVA/Padova simulator to adjust and validate four insulin-plus-pramlintide control algorithms. The proposals are based on an existing insulin controller and administer pramlintide either as independent boluses or as a ratio of the insulin infusion. The results of the insulin-pramlintide algorithms are compared against their insulin-alone counterparts, showing an improvement in the time in range between 3.00\% and 10.53\%, consistent with results reported in clinical trials in the literature. Future work will focus on individualizing the pramlintide model \rev{to the patients' characteristics} and evaluating the implemented strategies under more challenging scenarios.\\
\end{abstract}


%

\begin{keyword}
artificial pancreas\sep insulin-plus-pramlintide strategies\sep pramlintide PK/PD model\sep in silico testing\sep type 1 diabetes mellitus



\end{keyword}

\end{frontmatter}



\onecolumn

\noindent\fbox{%
	\parbox{\textwidth}{%
This article was published in Computers in Biology and Medicine. See \url{https://doi.org/10.1016/j.compbiomed.2025.110447} to access the editorial version of the article.
	}%
}

\twocolumn

\section{Introduction}

Type 1 Diabetes Mellitus (T1DM) is a chronic disease characterized by the autoimmune destruction of the pancreatic $\beta$-cells responsible for producing different hormones, including insulin and amylin. Amylin has several functions in the organism, such as slowing gastric emptying, managing energy expenditure, inducing satiation, and inhibiting glucagon production \cite{Muller2017}. 

Pramlintide is an amylin analog that has been approved as adjunctive therapy for T1DM in the USA \cite{Lutz2022} and has been successful in delaying glucose excursions after a meal \citep{Kolterman1996,Woerle2008,Hinshaw2016}. Pramlintide's effectivity has motivated interest in \rev{its use in artificial pancreas systems since  }
one of the main contributions of pramlintide is alleviating or even removing carbohydrate counting \cite{Tsoukas2021a, Tsoukas2021}, which is one of the main challenges for artificial pancreas systems. However, pramlintide administration modifies glucose dynamics, which must be considered to avoid insulin overdosing that could lead to hypoglycemia \citep{Infante2021}.

\rev{Some clinical trials have tested the efficacy of using pramlintide alongside automatic insulin delivery systems. The systematic review by Torres-Castaño \citep{Torres-Castano2022} illustrated how only three clinical trials up to June 2021 had tested the use of an insulin-pramlintide co-formulation with artificial pancreas systems. First, in \citep{Haidar2020}, the authors tested the co-administration of both hormones in a 6 {\microg}/U ratio, achieving +10\% time in range improvement when comparing a rapid-insulin-only system to the rapid insulin and pramlintide setup.
Similarly, their subsequent works tested different insulin/pramlintide ratios (5~{\microg}/U \citep{Tsoukas2021} and 10~{\microg}/U \citep{Tsoukas2021a}), aiming to test the feasibility of the co-formulation in removing meal announcements. The overall results from the trials showed that using the co-formulation with simple meal announcements achieved similar results to the insulin-only system with full carbohydrate announcements. The latest work on this matter was published in \citep{Cohen2024}, where the tested ratio was 10~{\microg}/U, achieving a moderate improvement in time in range using pramlintide with simple meal announcements compared to the hybrid insulin-only system. 
Other works in the literature have tested the subcutaneous administration of pramlintide as boluses at mealtimes. The earliest one dates back to 2012 \citep{Weinzimer2012}, where 30-{\microg} pramlintide boluses were administered alongside meals while using an automatic insulin delivery system. The system with pramlintide also improved time in range compared to the use of insulin alone. The same strategy was followed in \citep{Renukuntla2014}, while other works like \citep{Sherr2016} opted for 60-{\microg} boluses. In both cases, the system with pramlintide outperformed the insulin-only time in range, with a 4\% and a 10\% improvement, respectively.}

Developing successful \rev{dual-hormone artificial pancreas systems} requires designing control algorithms that consider both insulin and pramlintide. An inexpensive way of evaluating these controllers is using computer simulations since in~silico testing shortens the development process compared to in vivo preclinical trials \citep{Wilinska2010}. In addition, simulators guarantee the reproducibility of the evaluations and allow testing of the controllers in high-risk scenarios that would be unfeasible in clinical trials \citep{Vettoretti2018}. Several simulators exist for developing insulin-alone or even insulin-glucagon controllers, for instance, the UVA/Padova simulator, which received the Food and Drug Administration approval to be a substitute for preclinical trials with animals \citep{DallaMan2014}, the Cambridge simulator \citep{Wilinska2010}, or the OHSU simulator \citep{Resalat2019}. Regarding the in silico testing of insulin-pramlintide controllers, although a few pharmacokinetics/pharmacodynamics (PK/PD) models have been proposed in the literature \citep{Fang2013,Ramkissoon2014,Micheletto2013,Furio-Novejarque2023}, \rev{they have not been used in the context of artificial pancreas systems} \citep{Micheletto2013,Samaroo2018}.

This work aims to evaluate insulin-plus-pramlintide dosing strategies in an in silico setting. In order to do so, a recent pramlintide PK/PD model \citep{Furio-Novejarque2023} was implemented into the UVA/Padova T1DM simulator \citep{DallaMan2014}. Four pramlintide strategies are built alongside the insulin-alone controller developed by \citep{Sanz2023}. The implemented strategies explore using pramlintide being co-administered along with insulin either as a fixed pramlintide/insulin ratio, emulating insulin-pramlintide co-formulations under research \cite{Andersen2024}, or independent pramlintide boluses.

The rest of the paper is organized as follows: Section~\ref{sec:methods} presents the Methods, describing first the pramlintide model and the T1DM simulator used for the in silico validations, as well as the main insulin controller and the proposed insulin-plus-pramlintide administration techniques, including their tuning and the in silico validation. Section~\ref{sec:results} includes the tuning results and assesses the results of the in silico validation, comparing each strategy with the insulin-alone controllers. Section~\ref{sec:discussion} lays out a discussion related to the results, and Section~\ref{sec:conclusions} presents the conclusions of this work.

\section{Methods}\label{sec:methods}

This section presents the methodology used to extend a T1DM simulator and evaluate insulin-plus-pramlintide dosing strategies in a \rev{simulation environment}. First, the modification of the simulator to include the pramlintide PK/PD model is described. Then, the main controller used for insulin administration is presented, followed by a description of four proposals for pramlintide delivery. Finally, the tuning process of the strategies and the in silico validations are introduced.

\subsection{Pramlintide PK/PD model and T1DM Simulator}

In this work, the meal model of the UVA/Padova simulator has been expanded with the pramlintide PK/PD model presented in \citep{Furio-Novejarque2023}. The following equations define the pramlintide model:

\begin{subequations}
	\begin{align}
		\dv{{Q}_1(t)}{t} &= a_s \cdot p(t) - (k\sub{q1} + k_{q12}) \cdot Q_1(t) \label{eq:SC1} \\
		\dv{{Q}_2(t)}{t} &= k\sub{q12} \cdot Q_1(t) - k\sub{q2} \cdot Q_2(t)\label{eq:SC2}\\ 
		\dv{{P}_1(t)}{t} &= \left(k\sub{q1}\cdot Q_1(t) + k\sub{q2} \cdot Q_2(t)\right) - k_{e} \cdot P_1(t) \label{eq:IV} \\ 
		P(t) &= \frac{P_1(t)}{V_P}\label{eq:P} \\   
		\dv{{P}\sub{eff}(t)}{t} & = k_a \cdot \left(P_1(t) - P\sub{eff}(t)\right) \label{eq:PD1}\\ 
		h(P\sub{eff}) &= \frac{n \cdot P\sub{eff}^{e}}{d^{e} + P\sub{eff}^{e}} \label{eq:PD2} \\
    \eta(P\sub{eff}) &= \frac{1}{1 + h(P\sub{eff})} \label{eq:eta}
	\end{align}
\end{subequations}

\noindent where $p(t)$ is the pramlintide administrated subcutaneously. The variables $Q_1 (t)$ and $Q_2 (t)$ are the two compartments describing subcutaneous pramlintide kinetics. $P_1 (t)$ represents the plasma compartment, and $P(t)$ is the plasma pramlintide concentration. An extra compartment, $P\sub{eff}(t)$, represents the pramlintide effect, which is later modulated by the function $h(P\sub{eff})$. Finally, $\eta(P\sub{eff})$ is a time-varying factor in the interval 0 to 1 that will produce a larger delay in gastric emptying the higher the concentration of pramlintide is.

The original meal model implemented in the UVA/Padova simulator was developed by Dalla Man \citep{DallaMan2006}. In this model, the glucose rate of appearance follows the three-compartment system described below:

\begin{subequations}\label{eq:meal}
	\begin{align}
		\dv{Q\sub{sto1}(t)}{t} &= U_g(t) - k\sub{g21} \cdot Q\sub{sto1}(t) \label{eq:Qsto1} \\ 
		\dv{Q\sub{sto2}(t)}{t} &= k\sub{g21} \cdot Q\sub{sto1}(t) - k\sub{empt}(Q_{sto}) \cdot Q\sub{sto2}(t) \label{eq:Qsto2} \\
		\dv{Q\sub{gut}(t)}{t}  &= k\sub{empt}(Q\sub{sto}) \cdot Q\sub{sto2}(t) - k\sub{abs} \cdot Q\sub{gut}(t) \label{eq:Qgut}
	\end{align}

	\noindent where $Q_{sto}(t)$ and $k_{empt}(Q_{sto})$ are defined as:

	\begin{align}
		Q_{sto}(t) =& Q_{sto1}(t) + Q_{sto2}(t) \label{eq:Qsto} \\
		k\sub{empt}(Q\sub{sto}) =& k\sub{min} + \frac{k\sub{max} - k\sub{min}}{2} \cdot \nonumber \\ 
		& \{ \tanh \left(\alpha \cdot (Q\sub{sto}(t) - b \cdot D)\right) \nonumber \\
		 - & \tanh \left(\beta \cdot (Q\sub{sto}(t) - c \cdot D)\right)+ 2 \} \label{eq:kempt} 
	\end{align}

  \noindent with $\alpha$ and $\beta$ defined as:

  \begin{align}
    \alpha = \frac{5}{2 \cdot D (1 - b)}, \quad \beta = \frac{5}{2Dc}
  \end{align}

	\noindent with $D$ the amount of carbohydrates ingested. The meal rate of glucose appearance $(R_a(t))$, which is the total contribution of the meal to the amount of plasma glucose concentration, is defined as follows:

	\begin{equation}
		R_a(t) = \frac{f \cdot k\sub{abs} \cdot Q\sub{gut}(t)}{BW}
	\end{equation}
\end{subequations}

To incorporate the pramlintide effect in the UVA/Padova T1DM simulator, the gastric emptying rate $k\sub{empt}(Q\sub{sto})$ in equations \eqref{eq:Qsto2} and \eqref{eq:Qgut} has been multiplied by the attenuating factor $\eta(P\sub{eff})$ in \eqref{eq:eta}. Thus, the updated equations are defined as:

\begin{subequations}
	\begin{align}
		\dv{Q\sub{sto2}(t)}{t} &= k\sub{g21} \cdot Q\sub{sto1}(t) - \eta(P\sub{eff}) \cdot k\sub{empt}(Q\sub{sto}) \cdot Q\sub{sto2}(t) \\
		\dv{Q\sub{gut}(t)}{t}  &= \eta(P\sub{eff}) \cdot k\sub{empt}(Q\sub{sto}) \cdot Q\sub{sto2}(t) - k\sub{abs} \cdot Q\sub{gut}(t) 
	\end{align}
\end{subequations}

Table \ref{tb:parameters} contains the description and units of the parameters and states included in both the pramlintide PK/PD and the meal models' equations listed in this section. \rev{The model was identified using average aggregated data from the literature. The subcutaneous and intravenous PK stages used pramlintide dose administrations as input, and the output was compared to plasma pramlintide data. For the PD stage, the rate of glucose appearance data was used. More details about the identification process can be found in \citep{Furio-Novejarque2023}. Given that only average data was used to validate the model, the proposed model only comprises one set of parameters. Hence, all the virtual patients from the UVA/Padova simulator share the same pramlintide model.}


\begin{table*}[h]
	\caption{Pramlintide PK/PD and meal models parameters' descriptions and units.}\label{tb:parameters}
	\begin{tabular*}{\textwidth}{ c  c  c }  
		\toprule
		Symbol & Units & Description \\
		\midrule
		$p(t)$ & pmol/min & Pramlintide subcutaneous infusion \\
		$Q_1(t)$ & pmol & First subcutaneous compartment \\
		$Q_2(t)$ & pmol & Second subcutaneous compartment \\
		$P_1(t)$ & pmol & Plasma pramlintide compartment \\
		$P(t)$ & pmol/L & Plasma pramlintide volume \\
		$P\sub{eff}(t)$ & pmol & Pramlintide effect compartment \\ \midrule
		$a_s$ &   -            & Pramlintide bioavailability \\
		$k_{q1}$ & min$^{-1}$  & Rate from first subcutaneous compartment to plasma \\
		$k_{q12}$ & min$^{-1}$ & Rate from first to second subcutaneous compartment \\
		$k_{q2}$ & min$^{-1}$  & Rate from second compartment to plasma \\
		$k_e$ &   min$^{-1}$   & Output rate from plasma compartment \\
		$V_P$ &   L            & Plasma distribution volume \\
		$k_a$ &  min$^{-1}$    & Rate in the pramlintide effect compartment \\
		$n$ &     -            & Numerator coefficient in Hill equation from $\eta(\mathcal{P})$ \\
		$d$ &     pmol         & Denominator in Hill equation from $\eta(\mathcal{P})$ \\
		$e$ &     -            & Exponent in Hill equation from $\eta(\mathcal{P})$\\
		\midrule
		$U_g(t)$ & mg/min 		& Meal input rate \\
		$Q\sub{sto1}(t)$ & mg 	& Solid phase of glucose in the stomach \\
		$Q\sub{sto2}(t)$ & mg 	& Liquid phase of glucose in the stomach \\
		$Q\sub{gut}(t)$ & mg 		& Glucose mass in the intestine \\
		$k\sub{empt}(Q\sub{sto})$ & min$^{-1}$ & Rate constant of gastric emptying \\ 
		$R_a(t)$ & mg/kg/min    & Glucose rate of appearance in plasma \\
		$D$ & mg          			&  Amount of ingested glucose \\ 
		\midrule
		$f$ & - &  Fraction of intestinal absorption that appears in plasma \\
		$BW$ & kg 			    & Body weight \\
		$\alpha$ & - & $k\sub{empt}$ decrease rate \\
		$\beta$ & - & $k\sub{empt}$ increase rate \\
		$b$  & $\%$ 		    & Percentage of the dose for which $k\sub{empt}$ decreases to $(k\sub{max} - k\sub{min})/2$\\
		$c$ & $\%$ 		    & Percentage of the dose for which $k\sub{empt}$ is back to $(k\sub{max} - k\sub{min})/2$\\
		$k\sub{abs}$ & min$^{-1}$  & Rate constant of intestinal absorption \\
		$k\sub{min}$ & min$^{-1}$  & $k\sub{empt}$ minimum value \\
		$k\sub{max}$ & min$^{-1}$  & $k\sub{empt}$ maximum value \\
		$k\sub{g21}$ & min$^{-1}$  & Grinding rate \\
		\bottomrule
	\end{tabular*}
\end{table*}

\subsection{Main controller}
\label{sec:controller}

The pramlintide administration strategies evaluated in this work have been incorporated into the insulin-alone controller developed recently by our group \citep{Sanz2023,Sanz2024}. This controller has been selected because it can operate either with meal announcements or without, achieving, in both cases, competitive results compared to the literature. Note, however, that the implemented pramlintide strategies do not depend on the insulin-alone controller architecture; hence, another controller could have been used.

The controller in \citep{Sanz2023}, framed in the disturbance-observer-based control technique, was designed to minimize the postprandial glucose peak after an unannounced meal while guaranteeing that a positive control action is calculated. Its equations in implicit form are described below:

\begin{subequations}
	\begin{align}
		\hat{d}(s) & = F(s) \cdot G_d(s)^{-1} \left[y(s) + G_u(s) \cdot u^I (s)\right] \\
		u^I(s) & = \gamma \cdot \hat{d}(s) \\
		y(s) & = G(s) - \frac{G_b}{s}
	\end{align}
\end{subequations}

\noindent where $s$ represents the Laplace variable, $u^I (s)$ is the incremental insulin infusion rate concerning the steady state infusion $u_b$, $G(s)$ is the current blood glucose measured by a continuous glucose monitoring device (CGM), and $y(s)$ is the incremental blood glucose value concerning the basal glucose value $G_b$. 
Remark that the control action {$u^I (s)$} is proportional to the estimated meal disturbance $\hat{d}(s)$, being $\gamma$ a positive constant defining the controller gain. The filter $F(s)=\frac{1}{(\theta s + 1)^3}$  is used to estimate the disturbance, with $\theta$, the time constant of the filter. $G_d (s)$ and $G_u (s)$ are the transfer functions for the meal and the insulin glycemic effect. Both are defined as follows:

\begin{align}
	&G_i(s) = \tilde{G}_i(s) e^{-15s}, \quad	i \in \{d,u\} \\
	&\text{with} \quad   \tilde{G}_i = \frac{K_i^j}{(\tau_{1i}s + 1)^2(\tau_{2i}s+1)} \nonumber
\end{align}

\noindent where $\tau_{1i}$ and $\tau_{2i}$ are time constants (in minutes) and $K_i^j > 0$ is a patient-tailored gain for the $j$-patient, which were adjusted in \citep{Sanz2023} for the 10 virtual adults in the UVA/Padova T1DM simulator.

The controller also manages meal announcements by calculating the following prandial insulin bolus ($u^B$) when the user announces a meal to the system:

\begin{equation}
	u^B = \nu \cdot \frac{\widehat{CHO}}{CIR}
\end{equation}

\noindent where $\widehat{CHO}$ is the meal carbohydrate content (g) estimated by the user, $CIR$ is the carbohydrate-to-insulin ratio (g/U), and $\nu$ is an attenuating safety factor set to 0.8 \citep{Sanz2023}. The controller also implements the non-interacting mechanism described in \citep{Brosilow2002} to prevent hypoglycemic events caused by excessive insulin delivery after a prandial bolus \cite{Sanz2024}.

Then, in this work, the controller is responsible for administrating insulin in a basal-bolus manner. The pramlintide is added as an additional control action that is delivered based on specific conditions described in the next section. \rev{For clarity, Table 1 in the supplementary material of the manuscript contains a summary and description of the parameters related to the main controller and the pramlintide administration strategies.}

\subsection{Insulin-plus-pramlintide administration strategies}\label{sec:pram_strategies}

Throughout this section, four insulin-plus-pramlintide administration strategies will be explained. The first two aim to release users from the burden of meal announcements, the third strategy requires subjects to announce meals, and the last utilizes a simplified meal announcement. Guidance to tune these strategies is provided in Section~\ref{sec:tuning}. 

Two pramlintide administration methods employed in the literature were assessed: a fixed pramlintide bolus (Strategy~1 and Strategy~3) \cite{Weinzimer2012,Sherr2016} and a fixed pramlintide-insulin ratio (Strategy~2 and Strategy~4) \cite{Haidar2020,Tsoukas2021a,Tsoukas2021,Cohen2024}. Remark that the latter administration method imitates a co-formulation of insulin and pramlintide under development \cite{Andersen2024}.

Just for convenience, the input to the pramlintide PK/PD model, $p(k)$, is divided into bolus $p^B (k)$ and infusion $p^I (k)$ delivery. Then, $p(k)$ is denoted as:

\begin{equation}
	p(k) = p^B(k)\delta(k) + p^I(k)
\end{equation}

\noindent where the index $k \in \mathbb{Z}_{\geq0}$ denotes the sampling instants corresponding to a sampling time of 5 minutes, and $\delta(k)$ denotes the discrete Dirac function.

Henceforth, pramlintide doses are defined in {\microg}, so it is necessary to perform a unit conversion as the PK/PD pramlintide model is defined in pmol. Furthermore, insulin is always delivered by the controller in a basal-bolus manner so that the total insulin infusion can be defined as:

\begin{equation}
	u(k) = u^B(k)\delta(k) + u^I(k)+ u_b(k)
\end{equation}

\subsubsection{Strategy 1: Pramlintide $\lambda$-{\microg} bolus without meal announcements} 

In this strategy, a pramlintide dosing logic is designed to complement the insulin basal infusion of the main controller. This strategy does not consider meal announcements; hence, $u^B (k)=0$. In addition, this strategy will consider pramlintide bolus only, hence $p^I (k)=0$.

Meals are disturbances that affect glucose levels, generally causing a pronounced increase. The insulin-alone controller described in Section \ref{sec:controller} responds to these quick changes in glucose by increasing insulin administration. Hence, these sudden changes in the slope of glucose level and insulin delivery are used as indicators of the presence of a meal. In particular, the rate of change of both magnitudes between two consecutive time instants is calculated as:

\begin{align}
	m_u(k) &= \frac{u(k) - u(k-1)}{\Delta t}, \quad \forall k \geq 1 \\
	m_G(k) &= \frac{G(k) - G(k-1)}{\Delta t}, \quad \forall k \geq 1
\end{align}

\noindent where $m_u (k)$ and $m_G (k)$, respectively, denote insulin and glucose slopes, $u(k)$ is the current insulin value, and $G(k)$ is the current blood glucose level measured by the CGM.

Then, $p^B(k)$ is obtained as:

\begin{equation}
	p^B(k) = 
	\begin{cases}
		\lambda, \quad B_1(k) \wedge B_2(k) \\
		0, \quad \text{otherwise}
	\end{cases}
\end{equation}

\noindent where $\lambda$ represents the pramlintide bolus value in {\microg}. $B_1 (k)$ and $B_2 (k)$ are both conditions that must be achieved simultaneously ($\wedge$ refers to the logic operator AND) to deliver pramlintide. The following expressions define these conditions:

\begin{align}
	B_1(k) = & \left(m_u(k-1)>0\right) \wedge \left(m_u(k)<0\right) \label{eq:B1} \\
	B_2(k) = & \left(u(k)>z_1\cdot u_b\right) \wedge \left(m_G(k)>z_2\right) \wedge \left(\Delta k_p > z_3\right) \label{eq:B2}
\end{align}

\noindent where $\Delta k_p$ is the number of iterations between the current one and the last iteration in which pramlintide was delivered.

The conditions represented in \eqref{eq:B1} and \eqref{eq:B2} act as a meal detection strategy, restricting the administration of pramlintide so that it only happens after a meal intake. On the one hand, $B_1 (k)$ enables the detection of a peak of the insulin delivery signal if (a) the insulin delivery signal slope was positive in the previous iteration and (b) it is negative in the current one. On the other hand, $B_2 (k)$ contains a series of thresholds set for the variables to act as a safety mechanism and limit pramlintide administration to the cases when (a) the value of insulin delivery is at least over $z_1$ times its basal value, (b) the slope of glucose increase is over $z_2$, and (c) more than $z_3$ iterations ($5 \cdot z_3$ minutes) have passed since the last pramlintide bolus.  

\subsubsection{Strategy 2: Pramlintide/Insulin basal $\rho$-{\microg}/U ratio without meal announcements} 

This proposal mimics a co-formulation of insulin and pramlintide using a pramlintide/insulin ratio without meal announcements similar to \citep{Majdpour2021}, but only in a basal manner (hence, $p^B (k)=0$). Some preliminary results showed more hypoglycemic events when pramlintide was delivered as a ratio of insulin when large meals were announced to the controller, so these strategies were discarded.

Pramlintide is given as a ratio of $\rho$ {\microg} per unit of insulin delivered, so its value $p^I (k)$ is proportional to the insulin delivery signal $u(k)$ as follows:

\begin{equation}
	p^I(k) = \rho \cdot u(k)
\end{equation}

In this case, insulin boluses are not delivered since meals are not announced, so $u^B (k)=0$.

\subsubsection{Strategy 3: Pramlintide $\varphi$-{\microg} fixed bolus with meal announcements}

The controller is set to administer pramlintide as a single $\varphi$-{\microg} fixed bolus at the meal announcement time, whereas insulin is given in a basal-bolus manner. No pramlintide infusion is administrated in this strategy, resulting in $p^I (k)=0$. The pramlintide bolus is calculated by:

\begin{equation}
	p^B(k) = 
	\begin{cases}
		\varphi, \quad \widehat{CHO} (k) > 0 \\
		0, \quad \text{otherwise}
	\end{cases}
\end{equation}

\noindent where $\varphi$ is the value of the bolus in {\microg}, and $\widehat{CHO} (k)$ is the amount of carbohydrates estimated by the user in the current iteration.

This strategy is similar to the one \rev{used} by Sherr J. et al.~\citep{Sherr2016}. The difference is that pramlintide boluses are delivered at the onset of the meal, and meals are announced to the controller.

\subsubsection{Strategy 4: Pramlintide/Insulin basal and bolus \mbox{$\delta$-{\microg}} ratio with simplified meal announcements}\label{sec:pram_strategies:4}

Inspired by \cite{Tsoukas2021a}, in this strategy, both pramlintide infusion and pramlintide boluses are delivered as a ratio of $\delta$-{\microg} units of the infused insulin:
\begin{align}
	p^I(k) &= \delta \cdot (u^I(k)+u_b(k))\\
	p^B(k) &= \delta \cdot u^B(k)
\end{align}

In addition, the user only announces the mealtime (simplified meal announcement). The controller assumes a meal with 25~g of carbohydrates to calculate the prandial boluses. Snacks are not announced.

\subsection{Tuning of the pramlintide administration strategies}\label{sec:tuning}

For each strategy, the values of the corresponding bolus dosages (i.e., {$\lambda$}, {$\varphi$}) or ratios (i.e., {$\rho$}, {$\delta$}) have been selected among those reported in clinical trials. Hence, we have evaluated bolus dosages of 
15 {\microg} \citep{Chase2009}, 
30 {\microg} \citep{Kolterman1996,Woerle2008,Hinshaw2016,Chase2009,Colburn1996,Kong1998,Heptulla2005,Ahren2002,Renukuntla2014}, 
 45~{\microg} \citep{Heptulla2005}, 
 60~{\microg} \citep{Sherr2016,Kong1998}, 
 90~{\microg} \citep{Kong1998}, 
 100~{\microg} \citep{Kolterman1996,Colburn1996}, 
 and 300~{\microg} \citep{Kolterman1996,Colburn1996}, 
 and ratios of 3~{\microg}/U \citep{Majdpour2021}, 
 6~{\microg}/U \citep{Haidar2020}, 
 10~{\microg}/U \citep{Tsoukas2021, Majdpour2021}, 
 12~{\microg}/U \citep{Majdpour2021}, 
 and 15~{\microg}/U \citep{Majdpour2021}. 
 The parameters {$\lambda$}, {$\varphi$}, {$\delta$}, and {$\rho$} have been chosen to achieve the longest time in range (time percentage spent at glucose levels between 70 and 180 mg/dL), with the lowest time in hypoglycemia (time percentage spent at glucose levels under 70 mg/dL). Furthermore, since pramlintide is related to gastrointestinal side effects \citep{Edelman2007}, the amount of pramlintide delivered should be considered in the selection criteria. Thus, when two or more tunings behave similarly in terms of time in range and time in hypoglycemia, the one with the lowest daily pramlintide delivered has been selected.

 Simulations are run to choose the best bolus dosage or ratio using a scenario that includes a cohort of 10 virtual adults for 14 days. Subjects receive three daily random meals following a normal distribution with means 8:00~h, 13:00~h, 20:00~h, and a standard deviation of $\pm$~20 min. The meals contained a nominal value of 35, 50, and 85 g of carbohydrates, respectively. The content of each meal had a coefficient of variation of $\pm$~30\%.

 The simulations included sensor noise. In addition, the following sources of variability were considered. The nominal values of the meal absorption rate ($k_{abs}$) and carbohydrate bioavailability ($f$) variables of Equation (15) were modified per meal with a uniform distribution of $\pm$~30\% and $\pm$~10\%, respectively. Additionally, the nominal values of the parameters modeling the insulin pharmacokinetics in \citep{DallaMan2014} were modified, corresponding to a uniform distribution of $\pm$~30\% for every meal. The circadian variation in insulin sensitivity was represented by 24-hour sinusoidal fluctuation with random amplitude obeying a uniform $\pm$~30\% and random phase. Finally, a misestimation of the meal carbohydrate content in the meal announcements was implemented following the model described in \citep{Kawamura2015}.

 Besides the pramlintide bolus doses or ratios, Strategy~1 requires setting the thresholds $z_1$, $z_2$, and $z_3$. These parameters were tuned by exhaustive simulations using the same scenario described above.


\subsection{In silico validation}\label{sec:insilico_val}

Once the best tuning for each strategy was decided (see Section~\ref{sec:tuning_results}), the results of the adjusted insulin-plus-pramlintide strategies were compared in silico against the insulin-alone controllers: with meal announcement (MA), with no meal announcement (NMA), and the simplified meal announcement (SMA) described in Section~\ref{sec:pram_strategies:4}.

The simulations consider the 10 virtual adults of the simulator in a 14-day length scenario. Unlike the tuning scenario, meals were drawn from the distributions of mealtime and carbohydrate amounts described in \citep{Camerlingo2021}, representing real-life meal patterns. 
Up to seven meals can be included in a day: three main meals (breakfast, lunch, dinner) and a varying number of snacks (up to four), resulting in 168.58 (57.14)~g daily carbohydrates (median (standard deviation)). 
\rev{Table 2 in the Supplementary Material lists the generated pattern of meals for this scenario, including the meal type, time, and amount.}
Lastly, the simulations also include the different sources of variability considered in the tuning scenario (see Section~\ref{sec:tuning}): sensor noise, parametric variation of $k_{abs}$, $f$, insulin pharmacokinetics, circadian insulin sensitivity variation, and misestimation of the carbohydrate content. 

The metrics computed for each strategy were the time in range, time percentage spent at blood glucose levels under 54~mg/dL, time percentage spent at blood glucose levels under 70~mg/dL, time percentage spent at blood glucose levels above 180 mg/dL, time percentage spent at blood glucose levels above 250~mg/dL, and the total daily pramlintide and insulin delivery \citep{Battelino2019}. Moreover, since the time spent in hypoglycemia was almost negligible, hypoglycemia occurrence is also described by the metric LBGI (Low Blood Glucose Index) for the convenience of the statistical analysis. The HBGI (High Blood Glucose Index) is also used for completeness. LBGI and HBGI refer to the risk of hypoglycemia or hyperglycemia, respectively \citep{Kovatchev2019}. The lower the values, the lower the risk of either condition.

The insulin-plus-pramlintide administration techniques were compared by measuring the improvement regarding the insulin-alone controller in the cases of MA, SMA, and NMA. Paired mean differences were calculated by fitting a linear mixed-effect model for each metric, considering each strategy as a fixed factor and the subjects as a random effect. Confidence intervals at 95\% were also reported. Time in hypoglycemia was excluded from the statistical analysis since it was zero for most patients. Linear mixed-effects models were fitted with the \textsl{lme4} package \citep{Bates2015} of the statistical software R (version 3.4.1, \citep{R2021}). Lastly, paired mean differences were computed with the \textsl{marginaleffects} \cite{ArelXX} from the fitted linear mixed-effect models. A $p$-value lower than 0.05 was considered statistically significant. \text 

A positive value of this comparison in terms of time in range and negative values of the difference in time percentage spent at blood glucose levels either above 180 and 250~mg/dL (hyperglycemia) or under 70 and 54~mg/dL (hypoglycemia) mean that the pramlintide dosing strategy is providing better glycemic control than the insulin-alone operation modes. The same principle can be used regarding the LBGI and HBGI comparisons: negative values show that the corresponding pramlintide strategy achieves a reduced risk of hypoglycemia or hyperglycemia compared to its insulin-alone configurations.

\section{Results}\label{sec:results}

\subsection{Adjustment of $\lambda, \rho, \varphi$, and $\delta$ values}\label{sec:tuning_results}

\begin{figure}[t]
	\centering
	\includegraphics[width=\linewidth]{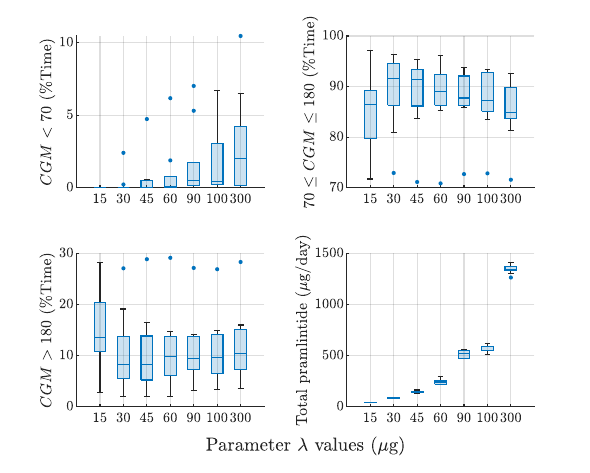}
	\caption{Results for the different values of $\lambda$ tested for the Strategy~1 tuning. The line inside each box represents the sample median, while the top and bottom edges are the 25th and 75th percentiles, respectively. Whiskers connect the upper or lower percentile to the nonoutlier maximum or minimum. Outliers, defined as values 1.5 times the interquartile range from the top or bottom box edges, are denoted by points.}
	\label{fig:tuning_strategy1}
\end{figure}

\begin{figure}[t]
	\centering
	\includegraphics[width=\linewidth]{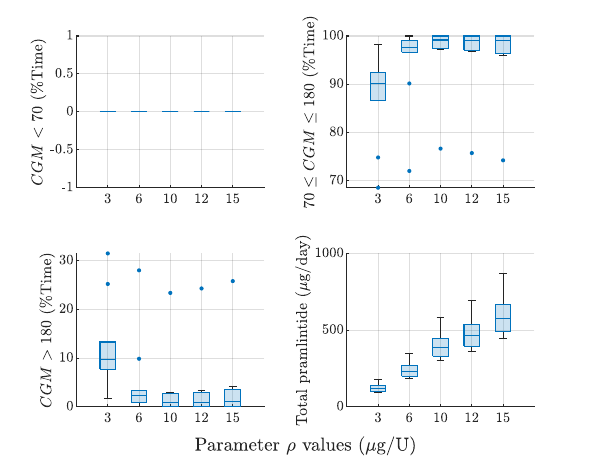}
	\caption{Results for the different values of $\rho$ tested for the Strategy~2 tuning. The line inside each box represents the sample median, while the top and bottom edges are the 25th and 75th percentiles, respectively. Whiskers connect the upper or lower percentile to the nonoutlier maximum or minimum. Outliers, defined as values 1.5 times the interquartile range from the top or bottom box edges, are denoted by points.}
	\label{fig:tuning_strategy2}
\end{figure}

\begin{figure}[t]
	\centering
	\includegraphics[width=\linewidth]{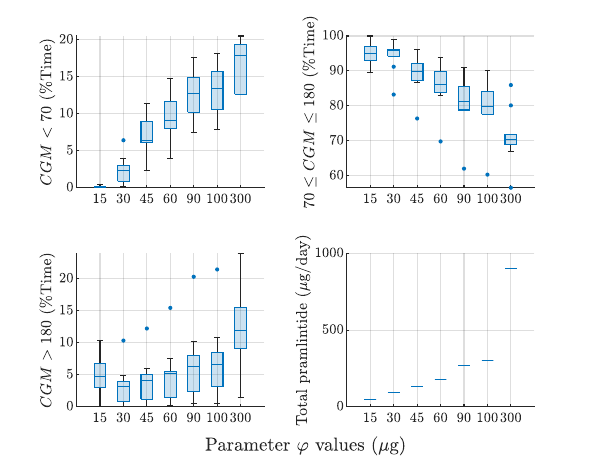}
	\caption{Results for the different values of $\varphi$ tested for the Strategy~3 tuning. The line inside each box represents the sample median, while the top and bottom edges are the 25th and 75th percentiles, respectively. Whiskers connect the upper or lower percentile to the nonoutlier maximum or minimum. Outliers, defined as values 1.5 times the interquartile range from the top or bottom box edges, are denoted by points.}
	\label{fig:tuning_strategy3}
\end{figure}

\begin{figure}[!h]
	\centering
	\includegraphics[width=\linewidth]{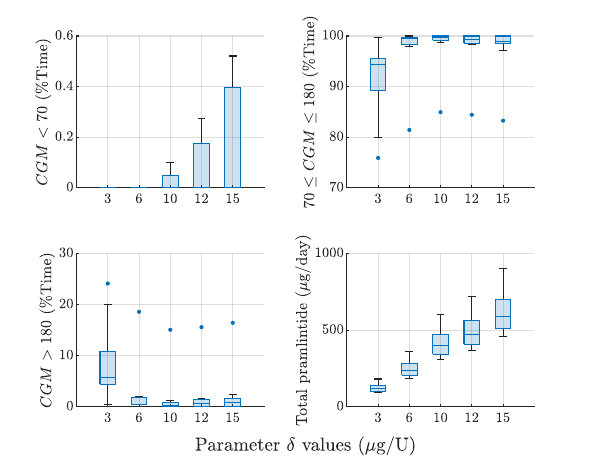}
	\caption{Results for the different values of $\delta$ tested for the Strategy~{4} tuning. The line inside each box represents the sample median, while the top and bottom edges are the 25th and 75th percentiles, respectively. Whiskers connect the upper or lower percentile to the nonoutlier maximum or minimum. Outliers, defined as values 1.5 times the interquartile range from the top or bottom box edges, are denoted by points.}
	\label{fig:tuning_strategy4}
\end{figure}

Figures~\ref{fig:tuning_strategy1} to~\ref{fig:tuning_strategy4} show the \rev{tuning} results obtained for each strategy described in Section~\ref{sec:pram_strategies}, employing the scenario defined in Section~\ref{sec:tuning}. Boxplots represent the overall metrics obtained for the 10 virtual adults in terms of time percentage spent at blood glucose levels between 70 and 180 mg/dL (time in range), above 180 mg/dL, below 70~mg/dL, and the total pramlintide delivered per day.

For Strategy 1, the best solution was found in setting $\lambda$ = 30 {\microg} since it achieved the most considerable time in range and the lowest time percentage at hypoglycemia (see Figure~\ref{fig:tuning_strategy1}). Note that higher doses of pramlintide increased the time percentage in hypoglycemia but barely changed the percentage in hyperglycemia.

Regarding Strategy 2, for which meals were not announced, the higher the ratio, the better results were achieved. The value of $\rho$ was eventually set to 10 {\microg}/U (see Figure~\ref{fig:tuning_strategy2}) as a trade-off tuning between the achieved time in range and the amount of pramlintide delivered.

The tuning results of Strategy~3 are similar to those of Strategy~2, the other strategy based on pramlintide boluses. There is a certain minimum amount for which a higher or a lower dose of pramlintide will not imply obtaining better results. Indeed, the time in range was virtually the same using either a fixed bolus of 30~{\microg} or 15~{\microg}. Nevertheless, the fewest hypoglycemic events were shown setting $\varphi$ = 15~{\microg} (see Figure~\ref{fig:tuning_strategy3}). 

For Strategy 4, a ratio of $\delta=10$~\microg/U maximizes the time in range while achieving an acceptable time in hypoglycemia (see Figure~\ref{fig:tuning_strategy4}).

\subsection{Study outcomes}
\rev{After running the insulin-plus-pramlintide strategies defined in Section~\ref{sec:pram_strategies} and the insulin-alone controller using the validation scenario (Section~\ref{sec:insilico_val}), results were evaluated by computing the metrics explained in the same section}. These metrics and the statistical analysis of the difference among strategies are summarized in Table~\ref{tb:descriptive} and Table~\ref{tb:contrasts}, respectively. 


{
	
	\begin{table*}[!t]
		\caption{Metrics of the insulin-plus-pramlintide administration strategies and the insulin-alone controller operation modes.}\label{tb:descriptive}
		 
		\begin{tabular*}{\linewidth}{@{\extracolsep{\fill}}lccccccc}
			\toprule
			\multirow{2}{*}{Metrics}&   \multicolumn{4}{c}{Pramlintide strategy} & \multicolumn{3}{c}{Insulin-Alone mode} \\ 
			\cmidrule(lr){2-5} \cmidrule(lr){6-8}
			&     1  & 2 & 3  & 4  & MA & SMA & NMA \\ 
			\midrule\addlinespace[2.5pt]
			\%Time $<$ 54 mg/dL (\%) &  0.0 (0.0) & 0.0 (0.0) & 0.0 (0.1) & 0.0 (0.0) & 0.0 (0.0) & 0.0 (0.0) & 0.0 (0.0) \\ 
			\%Time $<$ 70 mg/dL (\%) &  0.1 (0.3) & 0.0 (0.0) & 0.1 (0.3) & 0.1 (0.1) & 0.0 (0.0) & 0.0 (0.0) & 0.0 (0.0) \\ 
		    \%Time [70,180] mg/dL (\%) &  90.1 (9.4) & 97.6 (5.8) & 94.7 (5.3) & 99.2 (1.8) & 93.3 (6.0) & 90.3 (7.1) & 87.1 (8.5) \\ 
			\%Time $>$ 180 mg/dL (\%) &  9.8 (9.5) & 2.4 (5.8) & 5.2 (5.4) & 0.7 (1.9) & 6.6 (6.0) & 9.7 (7.1) & 12.9 (8.5) \\ 
			\%Time $>$ 250 mg/dL (\%) &  1.3 (1.8) & 0.0 (0.1) & 0.5 (0.9) & 0.0 (0.0) & 0.6 (1.1) & 1.4 (2.5) & 2.0 (3.3) \\ 
			LBGI &  0.1 (0.1) & 0.0 (0.0) & 0.2 (0.1) & 0.0 (0.0) & 0.2 (0.1) & 0.1 (0.1) & 0.1 (0.0) \\ 
			HBGI &  2.6 (1.8) & 1.7 (1.3) & 1.6 (1.0) & 1.1 (1.0) & 1.8 (1.1) & 2.4 (1.5) & 3.1 (1.9) \\ 
			Daily insulin (U) &  39.9 (8.6) & 39.1 (8.3) & 41.8 (9.4) & 40.3 (8.8) & 41.4 (9.2) & 40.4 (8.8) & 39.5 (8.5) \\ 
			Daily pramlintide ({\microg}) &  83.6 (6.3) & 391.3 (83.1) & 62.1 (0.0) & 403.2 (87.6) & 0.0 (0.0) & 0.0 (0.0) & 0.0 (0.0) \\ 
			\bottomrule
		\end{tabular*}
		\begin{minipage}{\linewidth}
			The results are expressed in mean (standard deviation). Strategy 1, 2, 3, and 4 denote the pramlintide 30-{\microg} bolus strategy without meal announcements, the strategy using a pramlintide/insulin basal 10-{\microg}/U ratio without meal announcements, the pramlintide 15-{\microg} fixed bolus strategy with meal announcements and the strategy using a pramlintide/insulin basal 10-{\microg}/U ratio with simplified meal announcements, respectively. Of note, MA denotes Meal Announcement, NMA denotes No Meal Announcement, and SMA Simple Meal Announcement.\\
		\end{minipage}
	\end{table*}
	
}

{
	\setlength{\tabcolsep}{4pt}
\begin{table*}[!t]
	\caption{Comparison between the insulin-plus-pramlintide dosing strategies and the insulin-alone deliveries.}\label{tb:contrasts}
	\center
	\resizebox{\textwidth}{!}{
	\begin{tabular}{m{0.11\textwidth}>{\centering}m{0.08\textwidth}>{\centering}m{0.15\textwidth}>{\centering}m{0.15\textwidth}>{\centering}m{0.15\textwidth}>{\centering}m{0.15\textwidth}>{\centering\arraybackslash}m{0.15\textwidth}}
		\toprule
		&  & \multicolumn{5}{c}{Difference in means between comparator and strategy} \\ 
		\cmidrule(lr){3-7}
		Insulin-alone comparator & Praml. strategy & \%Time [70,180] mg/dL & \%Time $>$180 mg/dL & \%Time $>$250 mg/dL & LBGI & HBGI \\ 
		\midrule\addlinespace[2.5pt]
		\multirow{8}{*}{\parbox{1.2cm}{Insulin-Alone (MA)}} & \multirow{2}{*}{1} &  -3.17&   2.99 &   0.64 &  -0.05 &   0.76 \\
		& &   [-5.21,  -1.12]* &   [0.98,   4.99]* &  [-0.07,   1.35]  &  [-0.09,  -0.01]* &   [0.42,   1.10]* \\  
		& \multirow{2}{*}{2} &   4.35  &  -4.35 &  -0.37 &  -0.15  &  -0.10  \\
		& &   [2.16,   6.53]* &  [-6.57,  -2.13]* & [-0.75,   0.01]  & [-0.20,  -0.11]* &[-0.44,   0.23]  \\ 
		& \multirow{2}{*}{3} &   1.33&  -1.49&   0.00 &   0.02 &  -0.25\\ 
		& &   [-0.24,   2.90]  &   [-3.06,   0.09]  &   [-0.32,   0.33]  &    [-0.03,   0.06]  &   [-0.59,   0.09]  \\ 
		& \multirow{2}{*}{4} &   5.43&  -5.16 &  -0.37&  -0.11  &  -0.68 \\
		& &    [2.89,   7.98]* &  [-7.66,  -2.66]* &   [-0.76,   0.01]  &  [-0.15,  -0.07]* &  [-1.02,  -0.34]* \\  
		\midrule\addlinespace[2.5pt]
		\multirow{8}{*}{\parbox{1.2cm}{Insulin-Alone (SMA)}} & \multirow{2}{*}{1} &  -0.10  &  -0.10  &   0.04  &   0.00   &   0.14  \\
		&  &   [-1.95,   1.75]  &   [-1.93,   1.73]  &    [-0.64,   0.72]  &    [-0.04,   0.05]  &    [-0.20,   0.48]  \\  
		& \multirow{2}{*}{2} &   7.41 &  -7.44 &  -0.98  &  -0.10 &  -0.73  \\
		&  &   [4.20,  10.62]* &  [ -10.73,  -4.14]* &   [-1.89,  -0.06]* &  [-0.14,  -0.06]* &  [-1.07,  -0.39]* \\ 
		& \multirow{2}{*}{3} &   4.40 &  -4.57 &  -0.60  &   0.07  &  -0.88  \\
		& &   [2.10,   6.70]* &   [-6.94,  -2.20]* &   [-1.33,   0.12]  &   [0.03,   0.11]* &   [-1.21,  -0.54]* \\ 
		& \multirow{2}{*}{4} &   8.50  &  -8.25  &  -0.98  &  -0.06  &  -1.30 \\
		&  &   [4.89,  12.12]* &   [ -11.85,  -4.65]* &  [-1.90,  -0.06]* &  [-0.10,  -0.01]* &   [-1.64,  -0.96]* \\
		\midrule\addlinespace[2.5pt] 
		\multirow{8}{*}{\parbox{1.2cm}{Insulin-Alone (NMA)}} & \multirow{2}{*}{1} &   3.02  &  -3.22  &  -0.82&   0.03  &  -0.49 \\
		 &  &   [0.82,   5.22]* &   [-5.45,  -1.00]* & [-1.79,   0.14]  &   [-0.02,   0.07]  &  [-0.83,  -0.15]* \\ 
		& \multirow{2}{*}{2} &  10.53 & -10.56 &  -1.84 &  -0.08 &  -1.36 \\
		&  &   [6.35,  14.72]* & [ -14.86,  -6.26]* &  [-3.36,  -0.31]* &  [-0.12,  -0.03]* &  [-1.70,  -1.02]* \\  
		& \multirow{2}{*}{3} &   7.52  &  -7.70  &  -1.46  &   0.09  &  -1.51  \\
		&  &   [4.36,  10.68]* &   [ -10.97,  -4.42]* &   [-2.73,  -0.19]* &   [0.05,   0.14]* &   [-1.85,  -1.17]* \\ 
		& \multirow{2}{*}{4} &  11.62 & -11.37 &  -1.84 &  -0.03  &  -1.93 \\
		&  &  [7.01,  16.23]* &  [ -15.99,  -6.75]* &  [-3.36,  -0.32]* &   [-0.08,   0.01]  &  [-2.27,  -1.59]* \\  
		\bottomrule
	\end{tabular}
	}
	\begin{minipage}{\linewidth}
		The mean differences, Comparator minus Strategy, are shown with the 95\% confidence interval. Pramlintide delivery strategies are denoted by Strategy 1 (pramlintide 30-{\microg} bolus strategy without meal announcements), 2 (pramlintide/insulin basal 10-{\microg}/U ratio without meal announcements), 3 (pramlintide 15-{\microg} fixed bolus strategy with meal announcements), and 4 (pramlintide/insulin basal 10-{\microg}/U ratio with simplified meal announcements).
		Of note, for the insulin-alone strategies, MA denotes Meal Announcement, NMA denotes No Meal Announcement, and SMA simple meal announcement. The symbol * denotes a $p$-value \textless{} 0.05.\\
	\end{minipage}
\end{table*}
}


\subsubsection{Strategy 1: Pramlintide 30-{\microg} bolus without meal announcements}

According to the statistical comparisons in Table 3, this strategy increases the time in range compared to the insulin-alone controller without meal announcements (+3.02\%). It also reduces time above 180 mg/dL ($-$3.22\%) and HBGI ($-$0.49). \rev{This leads to a time in range of 90.1\% ($\pm$9.4), compared to the 87.1\% ($\pm$8.5) of the NMA insulin-alone mode (Table 2). According to Tables 2 and~3, barely any hypoglycemia is reported (LGBI of 0.03).}
Conversely, compared to the insulin-alone system with meal announcement, it lowers time in range ($-$3.17\%) and increases time percentage above 180 mg/dL (+2.99\%) and HBGI (+0.76).

Concerning hypoglycemia, Table~\ref{tb:descriptive} reports a time percentage spent at blood glucose levels under 70 mg/dL of only about 0.10\%. Furthermore, the LBGI is 0.05 lower than the index obtained with the insulin-alone mode with meal announcements (Table~\ref{tb:contrasts}).

\rev{Regarding the SMA insulin-alone mode, this strategy hardly affects the time in range (-0.10\%) or the time percentage above 180 mg/dL (-0.10\%), as can be seen in Table 3. Table 2 shows negligible metric differences between these two configurations.}

\subsubsection{Strategy 2: Pramlintide/Insulin basal 10-{\microg}/U ratio without meal announcements}

Using pramlintide as a basal infusion along with insulin improves the time in range and time above 180~mg/dL by 4.35\% compared to the insulin-alone configuration with meal announcements (Table~\ref{tb:contrasts}). As expected, this magnitude of improvement is more considerable than the NMA and SMA cases, with improvements of 10.50\% and 7.41\%, respectively. Indeed, regarding NMA and SMA, Strategy~2 achieved a significant reduction in time percentage above 250~mg/dL and the HBGI.

In terms of hypoglycemia, this strategy decreases the LBGI the most among strategies regarding all insulin-alone operation modes (Table~\ref{tb:contrasts}).
\rev{These results can be seen in Table \ref{tb:descriptive}, where the time in range, time above 180~mg/dL, and time below 70 mg/dL are improved when compared to all insulin-alone modes. Of note, it also presents a large amount of daily pramlintide administered (391.3~{\microg} on average), which is discussed further below.}

\subsubsection{Strategy 3: Pramlintide 15-{\microg} fixed bolus with meal announcements}

A fixed pramlintide prandial bolus of 15 {\microg} allows an improvement of 1.33\% in time in range compared to the insulin-alone MA. Table 3 also shows a 1.49\% reduction in time spent at blood glucose levels above 180 mg/dL. The HBGI is reduced by 0.25. In addition, there is no statistically significant reduction in the LBGI.

Compared to the insulin-alone NMA and SMA, Table~\ref{tb:contrasts} reports a 7.52\% and a 4.40\%. However, it should be recalled that the insulin-alone MA already implies an increase in time in range. In particular, following the same procedure described in Section~\ref{sec:insilico_val}, these improvements are 6.19 [3.39,8.99]\% and 3.07 [1.03, 5.11]\% regarding insulin-alone NMA and SMA, respectively. As a result, adding pramlintide improves the performance by about 1\%.


Regarding hypoglycemia, the percentage time in hypoglycemia is slightly larger (Table~\ref{tb:descriptive}). However, the increase in LBGI is limited, although statistically different from zero regarding the SMA and NMA cases (Table~\ref{tb:contrasts}).

\rev{Table \ref{tb:descriptive} shows an overall improvement as time in range, which is on average, 94.7\% ($\pm$5.3). The other metrics are lower than or on par with those of the three insulin-alone modes. }

\subsubsection{Strategy 4: Pramlintide/Insulin basal 10-{\microg}/U ratio with simplified meal announcements}

Adding pramlintide to the insulin-alone SMA leads to an improvement in all the metrics of Table~\ref{tb:descriptive}, especially for the time in range (+8.50\%) and time above 180~mg/dL (+8.25\%), \rev{as shown in Table \ref{tb:contrasts}}. Strategy~4 also increases the time in range (+5.43\%) and reduces time above 180~mg/dL ($-$5.16\%) regarding the insulin-alone MA configuration. \rev{Compared to the insulin-alone NMA mode in Table 3, this strategy shows the greatest improvement in terms of time in range (+11.62\%) and time above 180~mg/dL (-11.37\%) amongst all strategies. }
Furthermore, Table~\ref{tb:descriptive} does not show a significant change for the LBGI either. Indeed, this outcome tends to decrease compared to all insulin-alone configurations (Table~\ref{tb:contrasts}).

\section{Discussion} \label{sec:discussion}
In this work, four pramlintide administration strategies were tuned and validated with the UVA/Padova simulator \cite{DallaMan2014}. This simulator was extended with a pramlintide PK/PD model \cite{Furio-Novejarque2023} to enable in silico validations of \rev{insulin-plus-pramlintide control algorithms}. The four implemented strategies were built upon the insulin-alone controller described in \cite{Sanz2023, Sanz2024}. Two of the strategies were based on the meal-announcement-free configuration of the controller (Strategy 1 and Strategy 2), while Strategy~3 considers full meal announcement, and Strategy~4 assumes a simplified meal announcement. The implemented strategies reproduce some ways of administrating pramlintide found in the literature, such as pramlintide boluses or pramlintide infusions calculated as a ratio of insulin delivery.

\begin{figure}[h]
	\centering
	\includegraphics[width=\linewidth]{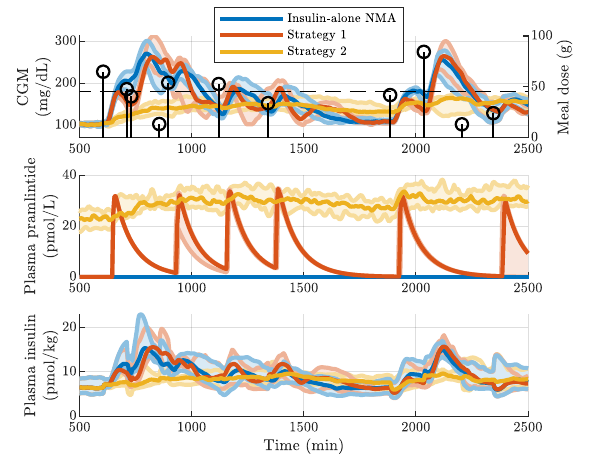}
	\caption{Overall comparison of Strategy 1, Strategy 2, and the insulin-alone controller without meal announcements (NMA) for the 10 virtual adults in the simulator. The top plot shows CGM values with meal carbohydrate content represented with a circle. The dashed line indicates the hyperglycemia threshold (180 mg/dL). The middle plot shows plasma pramlintide, and the bottom plot shows plasma insulin. Thick lines correspond to the median values of the ten virtual adults, and shaded areas represent the 25th and 75th percentiles.}
	\label{fig:strategy1vsstrategy2}
\end{figure}

Adding pramlintide to the corresponding base main controller enhances the pramlintide-free counterpart by increasing the time in range percentage and reducing the time in hyperglycemia percentage. However, the comparisons with the meal-announcement-free strategies (Strategy 1 and Strategy 2) regarding the insulin-alone controller with insulin prandial boluses show inconclusive results. The results achieved by Strategy 2 support that pramlintide may enable the removal of insulin prandial boluses, a conclusion that aligns with \cite{Tsoukas2021}. Nevertheless, Strategy~1 still leads to statistically significantly higher time percentage in hyperglycemia and lower time in range than the insulin-alone controller with meal announcements. The differences in the pramlintide administration method may explain this discrepancy. Strategy~1 delivered the drug based on a meal-detection-like logic, so pramlintide is never administrated simultaneously with the meal. Figure~\ref{fig:strategy1vsstrategy2} shows how, in the first strategy, the CGM overlaps the insulin-alone controller mode without meal announcements until pramlintide is delivered and takes effect on the virtual patient (see orange and blue lines around time instants 620 and 1900~min).
Nevertheless, in Strategy 2, pramlintide is given as a continuous infusion; therefore, there is a certain amount of plasma pramlintide at mealtime. Hence, pramlintide can act right after its pharmacodynamic delay. This higher quantity allows for slowing gastric emptying in favor of a better glycemic control capability.

The improvements in time in range observed in the simulations with the pramlintide administration strategies are close to the ones reported in some clinical trials in the literature: Sherr J et al. \cite{Sherr2016}, which inspired Strategy~3, undertook a closed-loop clinical trial with T1DM adult patients delivering a 60-{\microg} pramlintide bolus 15 min before the meal and without insulin prandial boluses. They reported an improvement of 11.3\% in time in range, which is similar to the 7.52\% improvement reached by Strategy~3. Furthermore, the 10-{\microg/U} pramlintide-to-insulin ratio with simplified meal announcements evaluated in \cite{Tsoukas2021a} and extended in \cite{Cohen2024} outperformed, respectively, in 10\% and in 7\% (adult cohort only) the time in range of an insulin-alone system, also with simplified meal announcements. The simulation results of Strategy~4, based on \cite{Tsoukas2021a}, approximately reproduce these differences with an improvement of 8.50\% in the time in range compared to the insulin-alone counterpart. Both clinical trials also compared \rev{the insulin-pramlintide artificial pancreas} system with simplified meals and an insulin-alone system with carbohydrate counting. The insulin-pramlintide system with simplified meal announcements barely enhanced the outcomes of the insulin-alone system with carbohydrate counting in \cite{Tsoukas2021a}. However, the study in \cite{Cohen2024} resulted in a difference of 6\% in favor of the insulin-pramlintide system (although without concluding statistical superiority), which is close to the 5.43\% improvement in time in range achieved by Strategy~4 regarding the insulin-alone system with meal announcement (Table~\ref{tb:contrasts}). Even though similar results between the results of clinical trials and the simulations in this paper have been found, it should be pointed out that these comparisons are only qualitative since no head-to-head comparison is possible given the differences in insulin-plus-pramlintide administration manners and the scenarios used.

Since using high doses of pramlintide may cause gastrointestinal side effects \cite{Edelman2007}, the analysis of the strategies' performance also included the amount of delivered pramlintide. 
\rev{Strategies 2 and 4 administer at least 4.6 times more pramlintide than the first strategy and more than 6 times compared to the third one. This substantial increase in pramlintide dosage (especially around mealtimes) likely accounts for the superior glycemic control observed with these approaches.}
Strategy 2 (with a ratio of 10~\microg/U) delivered a daily pramlintide amount of 391.3~{\microg} on average. This amount is similar to the 396~{\microg}/day reported by Huffman D. et al. \cite{Huffman2009} but higher than the 320~{\microg}/day reported by Haidar A. et al. \cite{Haidar2020}. 
\rev{Nevertheless, the amounts are consistent with the total daily pramlintide administered in the trial by Andersen et al. \citep{Andersen2023}, where a 6 {\microg}/U pramlintide-insulin co-formulation was administered. The total pramlintide dosages varied approximately between 252 {\microg} and 486 {\microg} for the low- and high-insulin requirements groups, respectively. In the study, some adverse gastrointestinal effects were associated with the use of the co-formulation. The drug's long-term safety was further evaluated in \citep{Andersen2024}, where results showed that gastrointestinal adverse events decreased progressively over 4 months of treatment.} 

In order to avoid the potential adverse effects of pramlintide, some considerations to reduce the pramlintide delivery might include stopping pramlintide delivery during nighttime (since postprandial glycemic control is not usually needed) or even reducing the quantity around mealtimes to decrease the daily amount. Nevertheless, one of the limitations of using an insulin-pramlintide co-formulation is the impossibility of separating both compounds, forcing them to maintain the same ratio. On the other hand, a basal administration with a variable or an adjustable ratio would require using two pumps or a dual-chamber pump to manage the administration of both drugs separately. However, both options increase the burden of consumables for the patient. 

\begin{figure}[h]
	\centering
	\includegraphics[width=\linewidth]{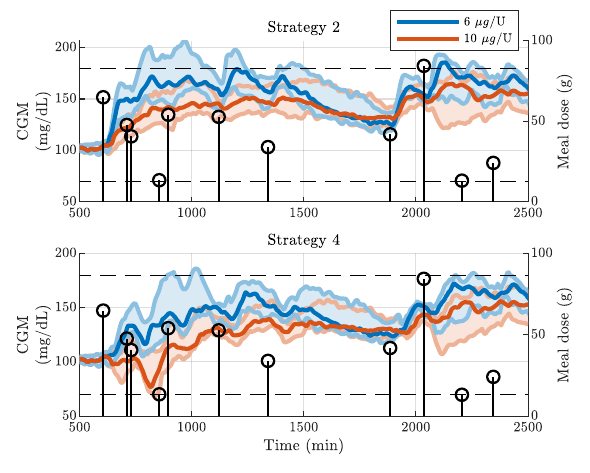}
	\caption{Overall comparison of the selected 10~\microg/U pramlintide-to-insulin ratio with the actual 6~\microg/U ratio under research for Strategy 2 (top panel) and Strategy 4 (bottom panel) and the 10 virtual adults in the simulator. Both panels show CGM values with meal carbohydrate content represented with a circle. Dashed lines indicate the hyperglycemia and hypoglycemia thresholds (180 and 70~mg/dL). Thick lines correspond to the median values of the ten virtual adults, and shaded areas represent the 25th and 75th percentiles.}
	\label{fig:6ugvs10ug}
\end{figure}

Regarding the pramlintide tuning, for the strategies that mimicked an insulin-pramlintide co-formulation (Strategy~2 and Strategy~4), the ratio that maximizes time in range was 10 \microg/U, a ratio already tested in clinical trials \cite{Tsoukas2021,Majdpour2021} using two independent insulin pumps. However, it should be noted that the actual insulin-pramlintide co-formulation under research utilizes a ratio of 6~\microg/U \cite{Andersen2023, Andersen2024}. To assess how using a 6-\microg/U ratio for Strategies 2 and 4 will impact their performance, we have performed new simulations with this ratio. The results indicate that considering a ratio of 6~\microg/U does not modify the overall conclusions: co-administring insulin and pramlintide enhances the performance achieved by the insulin-alone system counterpart, although the improvement is more modest than with the ratio of 10~\microg/U. In particular, using a 6-\microg/U ratio decreases the time in range achieved by the corresponding strategy with a ratio of 10~\microg/U in {2.17 [0.48, 3.8]\%} for Strategy 2 and 1.12 [3.39·10\textsuperscript{-3}, 2.27]\% for Strategy 4 (estimated difference in mean, [95\% confidence interval]). The reduction in time in range is due to a slight increase in time in hyperglycemia, as observed in Figure~\ref{fig:6ugvs10ug}. Conversely, an advantage of the ratio of 6~\microg/U is that it requires notably less pramlintide (Strategy~2 with a ratio 6~\microg/U: {236.2 (50.4)~\microg/day}, Strategy~4 with a ratio 6~\microg/U: 243.0~(53.6)~\microg/day) than the consumptions shown in Table~\ref{tb:descriptive}.       

Also related to the tuning, as observed from Figures~\ref{fig:tuning_strategy1}--\ref{fig:tuning_strategy4}, the strategies with meal announcements provided the lowest time in range at increasing pramlintide doses since the time percentage in hypoglycemia increased. The more considerable time in hypoglycemia could be due to the calculation of insulin prandial boluses, which did not account for the delayed glucose increase caused by the pramlintide administration. Future work will investigate techniques to modify the calculation of the insulin prandial bolus to consider the administration of pramlintide, such as attenuating the insulin bolus or delaying its administration.

Most of the limitations observed in this work are inherent to using a simulator and the simplifications needed. The same pramlintide model was used for each patient, so there were no differences among the patient’s reactions to the drug. However, patients may exhibit different tolerances to the drug in real life, modifying the required optimal pramlintide dosages. Another limitation is the simplicity of the scenario used, which only includes meal disturbances. Future work may incorporate physical activity events and investigating pramlintide interactions with rescue carbohydrates. It should be noted, however, that the simulation included variability in the meal model parameters and meal definition to bring the simulator closer to real-life conditions. 

\section{Conclusions}\label{sec:conclusions}
In this work, the UVA/Padova simulator, extended with a pramlintide model, has been utilized to adjust and validate four insulin-plus-pramlintide control algorithms. 
The implemented pramlintide administration algorithms were advantageous primarily compared to the insulin-alone mode of the controller since a higher time in range was reported without a remarkable increase in hypoglycemia. Improvements in terms of time in range were close to the improvements achieved in the literature. However,  given the differences in control algorithms and evaluation conditions, these results must be considered cautiously. \rev{The future scope of this project aims to investigate the pramlintide/insulin bolus dosing under different, more realistic scenarios (e.g., adding rescue carbohydrates and physical activity). Also, the pramlintide PK/PD model will be individualized for each virtual patient to account for patients’ inter-variability.} 

\section*{Declaration of interest}
The authors have nothing to declare.
\section*{Ethics statement}
The current study is focused on simulation, hence an statement of ethical approval is not required since no human subjects or animals were involved in the development of this work. 

\section*{CRediT authorship contribution statement}
\begin{creditlist}
	\credit{Borja Pons Torres}{Conceptualization, Formal analysis, Methodology, Software, Visualization, Writing – original draft, Writing – review \& editing}
	\credit{Iván Sala-Mira}{Conceptualization, Formal analysis, Methodology, Software, Supervision, Software, Visualization, Writing – original draft, Writing – review \& editing}
	\credit{Clara Furió-Novejarque}{Conceptualization, Resources, Supervision, Visualization, Writing – original draft, Writing – review \& editing}
	\credit{Ricardo Sanz}{Methodology, Resources, Supervision, Writing – review \& editing}
	\credit{Pedro García}{Supervision, Writing – review \& editing}
	\credit{José-Luis Díez}{Conceptualization, Funding acquisition, Project administration, Resources, Supervision, Writing – review \& editing}
	\credit{Jorge Bondia}{Conceptualization, Funding acquisition, Project administration, Resources, Supervision, Writing – review \& editing}
\end{creditlist}
\section*{Funding}

This work was supported by MCIN/AEI/10.13039/501100011033 under Grant PID2019-107722RB-C21; Conselleria de Innovacion, Universidades, Ciencia y Sociedad Digital from Generalitat Valenciana under Grant CIPROM/2021/012; and Plan de ayudas a la I+D+i del Instituto AI2, Convocatoria 2022.



  \bibliographystyle{elsarticle-num-names} 
  \bibliography{references}

\begin{thebibliography}{44}
\expandafter\ifx\csname natexlab\endcsname\relax\def\natexlab#1{#1}\fi
\providecommand{\url}[1]{\texttt{#1}}
\providecommand{\href}[2]{#2}
\providecommand{\path}[1]{#1}
\providecommand{\DOIprefix}{doi:}
\providecommand{\ArXivprefix}{arXiv:}
\providecommand{\URLprefix}{URL: }
\providecommand{\Pubmedprefix}{pmid:}
\providecommand{\doi}[1]{\href{http://dx.doi.org/#1}{\path{#1}}}
\providecommand{\Pubmed}[1]{\href{pmid:#1}{\path{#1}}}
\providecommand{\bibinfo}[2]{#2}
\ifx\xfnm\relax \def\xfnm[#1]{\unskip,\space#1}\fi
\bibitem[{M{\"{u}}ller et~al.(2017)M{\"{u}}ller, Finan, Clemmensen, {Di
  Marchi}, and Tsch{\"{o}}p}]{Muller2017}
\bibinfo{author}{T.~D. M{\"{u}}ller}, \bibinfo{author}{B.~Finan},
  \bibinfo{author}{C.~Clemmensen}, \bibinfo{author}{R.~D. {Di Marchi}},
  \bibinfo{author}{M.~H. Tsch{\"{o}}p},
\newblock \bibinfo{title}{{The new biology and pharmacology of glucagon}},
\newblock \bibinfo{journal}{Physiological Reviews} \bibinfo{volume}{97}
  (\bibinfo{year}{2017}) \bibinfo{pages}{721--766}.
  \DOIprefix\doi{10.1152/physrev.00025.2016}.
\bibitem[{Lutz(2022)}]{Lutz2022}
\bibinfo{author}{T.~A. Lutz},
\newblock \bibinfo{title}{{Creating the amylin story}},
\newblock \bibinfo{journal}{Appetite} \bibinfo{volume}{172}
  (\bibinfo{year}{2022}) \bibinfo{pages}{105965}.
  \DOIprefix\doi{10.1016/j.appet.2022.105965}.
\bibitem[{Kolterman et~al.(1996)Kolterman, Schwartz, Corder, Levy, Klaff,
  Peterson, and Gottlieb}]{Kolterman1996}
\bibinfo{author}{O.~G. Kolterman}, \bibinfo{author}{S.~Schwartz},
  \bibinfo{author}{C.~Corder}, \bibinfo{author}{B.~Levy},
  \bibinfo{author}{L.~Klaff}, \bibinfo{author}{J.~Peterson},
  \bibinfo{author}{A.~Gottlieb},
\newblock \bibinfo{title}{{Effect of 14 days' subcutaneous administration of
  the human amylin analogue, pramlintide (AC137), on an intravenous insulin
  challenge and response to a standard liquid meal in patients with IDDM}},
\newblock \bibinfo{journal}{Diabetologia} \bibinfo{volume}{39}
  (\bibinfo{year}{1996}) \bibinfo{pages}{492--499}.
  \DOIprefix\doi{10.1007/BF00400683}.
\bibitem[{Woerle et~al.(2008)Woerle, Albrecht, Linke, Zschau, Neumann,
  Nicolaus, Gerich, G{\"{o}}ke, and Schirra}]{Woerle2008}
\bibinfo{author}{H.~J. Woerle}, \bibinfo{author}{M.~Albrecht},
  \bibinfo{author}{R.~Linke}, \bibinfo{author}{S.~Zschau},
  \bibinfo{author}{C.~Neumann}, \bibinfo{author}{M.~Nicolaus},
  \bibinfo{author}{J.~E. Gerich}, \bibinfo{author}{B.~G{\"{o}}ke},
  \bibinfo{author}{J.~Schirra},
\newblock \bibinfo{title}{{Impaired hyperglycemia-induced delay in gastric
  emptying in patients with type 1 diabetes deficient for islet amyloid
  polypeptide}},
\newblock \bibinfo{journal}{Diabetes Care} \bibinfo{volume}{31}
  (\bibinfo{year}{2008}) \bibinfo{pages}{2325--2331}.
  \DOIprefix\doi{10.2337/dc07-2446}.
\bibitem[{Hinshaw et~al.(2016)Hinshaw, Schiavon, Dadlani, Mallad, {Dalla Man},
  Bharucha, Basu, Geske, Carter, Cobelli, Basu, and Kudva}]{Hinshaw2016}
\bibinfo{author}{L.~Hinshaw}, \bibinfo{author}{M.~Schiavon},
  \bibinfo{author}{V.~Dadlani}, \bibinfo{author}{A.~Mallad},
  \bibinfo{author}{C.~{Dalla Man}}, \bibinfo{author}{A.~Bharucha},
  \bibinfo{author}{R.~Basu}, \bibinfo{author}{J.~R. Geske},
  \bibinfo{author}{R.~E. Carter}, \bibinfo{author}{C.~Cobelli},
  \bibinfo{author}{A.~Basu}, \bibinfo{author}{Y.~C. Kudva},
\newblock \bibinfo{title}{{Effect of Pramlintide on Postprandial Glucose Fluxes
  in Type 1 Diabetes}},
\newblock \bibinfo{journal}{The Journal of Clinical Endocrinology and
  Metabolism} \bibinfo{volume}{101} (\bibinfo{year}{2016})
  \bibinfo{pages}{1954--1962}. \DOIprefix\doi{10.1210/jc.2015-3952}.
\bibitem[{Tsoukas et~al.(2021{\natexlab{a}})Tsoukas, Cohen, Legault, von
  Oettingen, Yale, Vallis, Odabassian, {El Fathi}, Rutkowski, Jafar, Ghanbari,
  Gouchie-Provencher, Ren{\'{e}}, Palisaitis, and Haidar}]{Tsoukas2021a}
\bibinfo{author}{M.~A. Tsoukas}, \bibinfo{author}{E.~Cohen},
  \bibinfo{author}{L.~Legault}, \bibinfo{author}{J.~E. von Oettingen},
  \bibinfo{author}{J.~F. Yale}, \bibinfo{author}{M.~Vallis},
  \bibinfo{author}{M.~Odabassian}, \bibinfo{author}{A.~{El Fathi}},
  \bibinfo{author}{J.~Rutkowski}, \bibinfo{author}{A.~Jafar},
  \bibinfo{author}{M.~Ghanbari}, \bibinfo{author}{N.~Gouchie-Provencher},
  \bibinfo{author}{J.~Ren{\'{e}}}, \bibinfo{author}{E.~Palisaitis},
  \bibinfo{author}{A.~Haidar},
\newblock \bibinfo{title}{{Alleviating carbohydrate counting with a
  FiASP-plus-pramlintide closed-loop delivery system (artificial pancreas):
  Feasibility and pilot studies}},
\newblock \bibinfo{journal}{Diabetes, Obesity and Metabolism}
  \bibinfo{volume}{23} (\bibinfo{year}{2021}{\natexlab{a}})
  \bibinfo{pages}{2090--2098}. \DOIprefix\doi{10.1111/dom.14447}.
\bibitem[{Tsoukas et~al.(2021{\natexlab{b}})Tsoukas, Majdpour, Yale, Fathi,
  Garfield, Rutkowski, Rene, Legault, and Haidar}]{Tsoukas2021}
\bibinfo{author}{M.~A. Tsoukas}, \bibinfo{author}{D.~Majdpour},
  \bibinfo{author}{J.~F. Yale}, \bibinfo{author}{A.~E. Fathi},
  \bibinfo{author}{N.~Garfield}, \bibinfo{author}{J.~Rutkowski},
  \bibinfo{author}{J.~Rene}, \bibinfo{author}{L.~Legault},
  \bibinfo{author}{A.~Haidar},
\newblock \bibinfo{title}{{A fully artificial pancreas versus a hybrid
  artificial pancreas for type 1 diabetes: a single-centre, open-label,
  randomised controlled, crossover, non-inferiority trial}},
\newblock \bibinfo{journal}{The Lancet Digital Health} \bibinfo{volume}{3}
  (\bibinfo{year}{2021}{\natexlab{b}}) \bibinfo{pages}{e723--e732}.
  \DOIprefix\doi{10.1016/S2589-7500(21)00139-4}.
\bibitem[{Infante et~al.(2021)Infante, Baidal, Rickels, Fabbri, Skyler,
  Alejandro, and Ricordi}]{Infante2021}
\bibinfo{author}{M.~Infante}, \bibinfo{author}{D.~A. Baidal},
  \bibinfo{author}{M.~R. Rickels}, \bibinfo{author}{A.~Fabbri},
  \bibinfo{author}{J.~S. Skyler}, \bibinfo{author}{R.~Alejandro},
  \bibinfo{author}{C.~Ricordi},
\newblock \bibinfo{title}{{Dual-hormone artificial pancreas for management of
  type 1 diabetes: Recent progress and future directions}},
\newblock \bibinfo{journal}{Artificial Organs} \bibinfo{volume}{45}
  (\bibinfo{year}{2021}) \bibinfo{pages}{968--986}.
  \DOIprefix\doi{10.1111/aor.14023}.
\bibitem[{Torres-Castaño et~al.(2022)Torres-Castaño, Rivero-Santana,
  Perestelo-Pérez, Duarte-Díaz, Abt-Sacks, Ramos-García, Álvarez Pérez,
  Wäagner, Rigla, and Serrano-Aguilar}]{Torres-Castano2022}
\bibinfo{author}{A.~Torres-Castaño}, \bibinfo{author}{A.~Rivero-Santana},
  \bibinfo{author}{L.~Perestelo-Pérez}, \bibinfo{author}{A.~Duarte-Díaz},
  \bibinfo{author}{A.~Abt-Sacks}, \bibinfo{author}{V.~Ramos-García},
  \bibinfo{author}{Y.~Álvarez Pérez}, \bibinfo{author}{A.~M. Wäagner},
  \bibinfo{author}{M.~Rigla}, \bibinfo{author}{P.~Serrano-Aguilar},
\newblock \bibinfo{title}{Dual-hormone insulin-and-pramlintide artificial
  pancreas for type 1 diabetes: A systematic review},
\newblock \bibinfo{journal}{Applied Sciences (Switzerland)}
  \bibinfo{volume}{12} (\bibinfo{year}{2022}) \bibinfo{pages}{1--15}.
  \DOIprefix\doi{10.3390/app122010262}.
\bibitem[{Haidar et~al.(2020)Haidar, Tsoukas, Bernier-Twardy, Yale, Rutkowski,
  Bossy, Pytka, {El Fathi}, Strauss, and Legault}]{Haidar2020}
\bibinfo{author}{A.~Haidar}, \bibinfo{author}{M.~A. Tsoukas},
  \bibinfo{author}{S.~Bernier-Twardy}, \bibinfo{author}{J.~F. Yale},
  \bibinfo{author}{J.~Rutkowski}, \bibinfo{author}{A.~Bossy},
  \bibinfo{author}{E.~Pytka}, \bibinfo{author}{A.~{El Fathi}},
  \bibinfo{author}{N.~Strauss}, \bibinfo{author}{L.~Legault},
\newblock \bibinfo{title}{{A novel dual-hormone insulin- and-pramlintide
  artificial pancreas for type 1 diabetes: A randomized controlled crossover
  trial}},
\newblock \bibinfo{journal}{Diabetes Care} \bibinfo{volume}{43}
  (\bibinfo{year}{2020}) \bibinfo{pages}{597--606}.
  \DOIprefix\doi{10.2337/dc19-1922}.
\bibitem[{Cohen et~al.(2024)Cohen, Tsoukas, Legault, Vallis, Oettingen,
  Palisaitis, Odabassian, Yale, Garfield, Gouchie-Provencher, Rutkowski, Jafar,
  Ghanbari, and Haidar}]{Cohen2024}
\bibinfo{author}{E.~Cohen}, \bibinfo{author}{M.~A. Tsoukas},
  \bibinfo{author}{L.~Legault}, \bibinfo{author}{M.~Vallis},
  \bibinfo{author}{J.~E.~V. Oettingen}, \bibinfo{author}{E.~Palisaitis},
  \bibinfo{author}{M.~Odabassian}, \bibinfo{author}{J.~F. Yale},
  \bibinfo{author}{N.~Garfield}, \bibinfo{author}{N.~Gouchie-Provencher},
  \bibinfo{author}{J.~Rutkowski}, \bibinfo{author}{A.~Jafar},
  \bibinfo{author}{M.~Ghanbari}, \bibinfo{author}{A.~Haidar},
\newblock \bibinfo{title}{Simple meal announcements and pramlintide delivery
  versus carbohydrate counting in type 1 diabetes with automated fast-acting
  insulin aspart delivery: a randomised crossover trial in montreal, canada},
\newblock \bibinfo{journal}{The Lancet Digital Health} \bibinfo{volume}{6}
  (\bibinfo{year}{2024}) \bibinfo{pages}{e489--e499}.
  \DOIprefix\doi{10.1016/S2589-7500(24)00092-X}.
\bibitem[{Weinzimer et~al.(2012)Weinzimer, Sherr, Cengiz, Kim, Ruiz, Carria,
  Voskanyan, Roy, and Tamborlane}]{Weinzimer2012}
\bibinfo{author}{S.~A. Weinzimer}, \bibinfo{author}{J.~L. Sherr},
  \bibinfo{author}{E.~Cengiz}, \bibinfo{author}{G.~Kim}, \bibinfo{author}{J.~L.
  Ruiz}, \bibinfo{author}{L.~Carria}, \bibinfo{author}{G.~Voskanyan},
  \bibinfo{author}{A.~Roy}, \bibinfo{author}{W.~V. Tamborlane},
\newblock \bibinfo{title}{Effect of pramlintide on prandial glycemic excursions
  during closed-loop control in adolescents and young adults with type 1
  diabetes},
\newblock \bibinfo{journal}{Diabetes care} \bibinfo{volume}{35}
  (\bibinfo{year}{2012}) \bibinfo{pages}{1994--1999}.
  \DOIprefix\doi{10.2337/DC12-0330}.
\bibitem[{Renukuntla et~al.(2014)Renukuntla, Ramchandani, Trast, Cantwell, and
  Heptulla}]{Renukuntla2014}
\bibinfo{author}{V.~S. Renukuntla}, \bibinfo{author}{N.~Ramchandani},
  \bibinfo{author}{J.~Trast}, \bibinfo{author}{M.~Cantwell},
  \bibinfo{author}{R.~A. Heptulla},
\newblock \bibinfo{title}{{Role of Glucagon-like peptide-1 analogue versus
  Amylin as an adjuvant therapy in type 1 diabetes in a closed loop setting
  with ePID algorithm}},
\newblock \bibinfo{journal}{Journal of Diabetes Science and Technology}
  \bibinfo{volume}{8} (\bibinfo{year}{2014}) \bibinfo{pages}{1011--1017}.
  \DOIprefix\doi{10.1177/1932296814542153}.
\bibitem[{Sherr et~al.(2016)Sherr, Patel, Michaud, Palau-Collazo, {Van Name},
  Tamborlane, Cengiz, Carria, Tichy, and Weinzimer}]{Sherr2016}
\bibinfo{author}{J.~L. Sherr}, \bibinfo{author}{N.~S. Patel},
  \bibinfo{author}{C.~I. Michaud}, \bibinfo{author}{M.~M. Palau-Collazo},
  \bibinfo{author}{M.~A. {Van Name}}, \bibinfo{author}{W.~V. Tamborlane},
  \bibinfo{author}{E.~Cengiz}, \bibinfo{author}{L.~R. Carria},
  \bibinfo{author}{E.~M. Tichy}, \bibinfo{author}{S.~A. Weinzimer},
\newblock \bibinfo{title}{{Mitigating meal-related glycemic excursions in an
  insulin-sparing manner during closed-loop insulin delivery: The beneficial
  effects of adjunctive pramlintide and liraglutide}},
\newblock \bibinfo{journal}{Diabetes Care} \bibinfo{volume}{39}
  (\bibinfo{year}{2016}) \bibinfo{pages}{1127--1134}.
  \DOIprefix\doi{10.2337/dc16-0089}.
\bibitem[{Wilinska et~al.(2010)Wilinska, Chassin, Acerini, Allen, Dunger, and
  Hovorka}]{Wilinska2010}
\bibinfo{author}{M.~E. Wilinska}, \bibinfo{author}{L.~J. Chassin},
  \bibinfo{author}{C.~L. Acerini}, \bibinfo{author}{J.~M. Allen},
  \bibinfo{author}{D.~B. Dunger}, \bibinfo{author}{R.~Hovorka},
\newblock \bibinfo{title}{{Simulation environment to evaluate closed-loop
  insulin delivery systems in type 1 diabetes}},
\newblock \bibinfo{journal}{Journal of Diabetes Science and Technology}
  \bibinfo{volume}{4} (\bibinfo{year}{2010}) \bibinfo{pages}{132--144}.
  \DOIprefix\doi{10.1177/193229681000400117}.
\bibitem[{Vettoretti et~al.(2018)Vettoretti, Facchinetti, Sparacino, and
  Cobelli}]{Vettoretti2018}
\bibinfo{author}{M.~Vettoretti}, \bibinfo{author}{A.~Facchinetti},
  \bibinfo{author}{G.~Sparacino}, \bibinfo{author}{C.~Cobelli},
\newblock \bibinfo{title}{{Type-1 diabetes patient decision simulator for in
  silico testing safety and effectiveness of insulin treatments}},
\newblock \bibinfo{journal}{IEEE Transactions on Biomedical Engineering}
  (\bibinfo{year}{2018}). \DOIprefix\doi{10.1109/TBME.2017.2746340}.
\bibitem[{{Dalla Man} et~al.(2014){Dalla Man}, Micheletto, Lv, Breton,
  Kovatchev, and Cobelli}]{DallaMan2014}
\bibinfo{author}{C.~{Dalla Man}}, \bibinfo{author}{F.~Micheletto},
  \bibinfo{author}{D.~Lv}, \bibinfo{author}{M.~Breton},
  \bibinfo{author}{B.~Kovatchev}, \bibinfo{author}{C.~Cobelli},
\newblock \bibinfo{title}{{The UVA/PADOVA type 1 diabetes simulator: New
  features}},
\newblock \bibinfo{journal}{Journal of Diabetes Science and Technology}
  \bibinfo{volume}{8} (\bibinfo{year}{2014}) \bibinfo{pages}{26--34}.
  \DOIprefix\doi{10.1177/1932296813514502}.
\bibitem[{Resalat et~al.(2019)Resalat, Youssef, Tyler, Castle, and
  Jacobs}]{Resalat2019}
\bibinfo{author}{N.~Resalat}, \bibinfo{author}{J.~E. Youssef},
  \bibinfo{author}{N.~Tyler}, \bibinfo{author}{J.~Castle},
  \bibinfo{author}{P.~G. Jacobs},
\newblock \bibinfo{title}{{A statistical virtual patient population for the
  glucoregulatory system in type 1 diabetes with integrated exercise model}},
\newblock \bibinfo{journal}{PLoS ONE} \bibinfo{volume}{14}
  (\bibinfo{year}{2019}) \bibinfo{pages}{e0217301}.
  \DOIprefix\doi{10.1371/journal.pone.0217301}.
\bibitem[{Fang et~al.(2013)Fang, Landersdorfer, Cirincione, and
  Jusko}]{Fang2013}
\bibinfo{author}{J.~Fang}, \bibinfo{author}{C.~B. Landersdorfer},
  \bibinfo{author}{B.~Cirincione}, \bibinfo{author}{W.~J. Jusko},
\newblock \bibinfo{title}{{Study Reanalysis Using a Mechanism-Based
  Pharmacokinetic/Pharmacodynamic Model of Pramlintide in Subjects with Type 1
  Diabetes}},
\newblock \bibinfo{journal}{The AAPS Journal} \bibinfo{volume}{15}
  (\bibinfo{year}{2013}) \bibinfo{pages}{15--29}.
  \DOIprefix\doi{10.1208/s12248-012-9409-7}.
\bibitem[{Ramkissoon et~al.(2014)Ramkissoon, Aufderheide, Bequette, and
  Palerm}]{Ramkissoon2014}
\bibinfo{author}{C.~M. Ramkissoon}, \bibinfo{author}{B.~Aufderheide},
  \bibinfo{author}{B.~W. Bequette}, \bibinfo{author}{C.~C. Palerm},
\newblock \bibinfo{title}{{A Model of Glucose-Insulin-Pramlintide
  Pharmacokinetics and Pharmacodynamics in Type I Diabetes}},
\newblock \bibinfo{journal}{Journal of Diabetes Science and Technology}
  \bibinfo{volume}{8} (\bibinfo{year}{2014}) \bibinfo{pages}{529--542}.
  \DOIprefix\doi{10.1177/1932296813517323}.
\bibitem[{Micheletto et~al.(2013)Micheletto, {Dalla Man}, Kolterman, Chiquette,
  Herrmann, Schirra, Kovatchev, and Cobelli}]{Micheletto2013}
\bibinfo{author}{F.~Micheletto}, \bibinfo{author}{C.~{Dalla Man}},
  \bibinfo{author}{O.~Kolterman}, \bibinfo{author}{E.~Chiquette},
  \bibinfo{author}{K.~Herrmann}, \bibinfo{author}{J.~Schirra},
  \bibinfo{author}{B.~Kovatchev}, \bibinfo{author}{C.~Cobelli},
\newblock \bibinfo{title}{{In silico design of optimal ratio for
  co-administration of pramlintide and insulin in type 1 diabetes}},
\newblock \bibinfo{journal}{Diabetes Technology and Therapeutics}
  \bibinfo{volume}{15} (\bibinfo{year}{2013}) \bibinfo{pages}{802--809}.
  \DOIprefix\doi{10.1089/dia.2013.0054}.
\bibitem[{Furi{\'{o}}-Novejarque et~al.(2024)Furi{\'{o}}-Novejarque, Sala-Mira,
  D{\'{i}}ez, and Bondia}]{Furio-Novejarque2023}
\bibinfo{author}{C.~Furi{\'{o}}-Novejarque}, \bibinfo{author}{I.~Sala-Mira},
  \bibinfo{author}{J.-L. D{\'{i}}ez}, \bibinfo{author}{J.~Bondia},
\newblock \bibinfo{title}{{A model of subcutaneous pramlintide pharmacokinetics
  and its effect on gastric emptying: Proof-of-concept based on populational
  data}},
\newblock \bibinfo{journal}{Computer Methods and Programs in Biomedicine}
  \bibinfo{volume}{244} (\bibinfo{year}{2024}) \bibinfo{pages}{107968}.
  \DOIprefix\doi{10.1016/j.cmpb.2023.107968}.
\bibitem[{Samaroo et~al.(2018)Samaroo, Wilkes, Ramkissoon, and
  Aufderheide}]{Samaroo2018}
\bibinfo{author}{A.~Samaroo}, \bibinfo{author}{M.~Wilkes},
  \bibinfo{author}{C.~Ramkissoon}, \bibinfo{author}{B.~Aufderheide},
\newblock \bibinfo{title}{{Optimal injection time of insulin to regulate
  postprandial glucose in type i diabetes in silico with pramlintide at
  meals}},
\newblock \bibinfo{journal}{Proceedings of the UTT Research Symposium}
  (\bibinfo{year}{2018}) \bibinfo{pages}{7--11}.
\bibitem[{Sanz et~al.(2023)Sanz, Garc{\'{i}}a, Romero-Viv{\'{o}}, D{\'{i}}ez,
  and Bondia}]{Sanz2023}
\bibinfo{author}{R.~Sanz}, \bibinfo{author}{P.~Garc{\'{i}}a},
  \bibinfo{author}{S.~Romero-Viv{\'{o}}}, \bibinfo{author}{J.~L. D{\'{i}}ez},
  \bibinfo{author}{J.~Bondia},
\newblock \bibinfo{title}{{Near-optimal feedback control for postprandial
  glucose regulation in type 1 diabetes}},
\newblock \bibinfo{journal}{ISA Transactions}  (\bibinfo{year}{2023}).
  \DOIprefix\doi{10.1016/j.isatra.2022.06.033}.
\bibitem[{Andersen et~al.(2024)Andersen, Eloy, Heise, Gaudier, Mégret,
  Seroussi, Chan, Soula, Riddle, and DeVries}]{Andersen2024}
\bibinfo{author}{G.~Andersen}, \bibinfo{author}{R.~Eloy},
  \bibinfo{author}{T.~Heise}, \bibinfo{author}{M.~Gaudier},
  \bibinfo{author}{C.~Mégret}, \bibinfo{author}{C.~Seroussi},
  \bibinfo{author}{Y.-P. Chan}, \bibinfo{author}{O.~Soula},
  \bibinfo{author}{M.~Riddle}, \bibinfo{author}{J.~H. DeVries},
\newblock \bibinfo{title}{Ado09, a co-formulation of pramlintide and insulin
  a21g, lowers body weight versus insulin lispro in type 1 diabetes},
\newblock \bibinfo{journal}{Diabetes, Obesity and Metabolism}
  \bibinfo{volume}{26} (\bibinfo{year}{2024}) \bibinfo{pages}{4639--4645}.
  \DOIprefix\doi{10.1111/dom.15827}.
\bibitem[{{Dalla Man} et~al.(2006){Dalla Man}, Camilleri, and
  Cobelli}]{DallaMan2006}
\bibinfo{author}{C.~{Dalla Man}}, \bibinfo{author}{M.~Camilleri},
  \bibinfo{author}{C.~Cobelli},
\newblock \bibinfo{title}{{A system model of oral glucose absorption:
  Validation on gold standard data}},
\newblock \bibinfo{journal}{IEEE Transactions on Biomedical Engineering}
  \bibinfo{volume}{53} (\bibinfo{year}{2006}) \bibinfo{pages}{2472--2478}.
  \DOIprefix\doi{10.1109/TBME.2006.883792}.
\bibitem[{Sanz et~al.(2024)Sanz, Sala-Mira, Furió-Novejarque, García, Díez,
  and Bondia}]{Sanz2024}
\bibinfo{author}{R.~Sanz}, \bibinfo{author}{I.~Sala-Mira},
  \bibinfo{author}{C.~Furió-Novejarque}, \bibinfo{author}{P.~García},
  \bibinfo{author}{J.-L. Díez}, \bibinfo{author}{J.~Bondia},
\newblock \bibinfo{title}{In silico validation of a customizable
  fully-autonomous artificial pancreas with coordinated insulin, glucagon and
  rescue carbohydrates},
\newblock \bibinfo{journal}{Biocybernetics and Biomedical Engineering}
  \bibinfo{volume}{44} (\bibinfo{year}{2024}) \bibinfo{pages}{560--568}.
  \DOIprefix\doi{10.1016/j.bbe.2024.08.003}.
\bibitem[{Brosilow and Babu(2002)}]{Brosilow2002}
\bibinfo{author}{C.~Brosilow}, \bibinfo{author}{J.~Babu},
  \bibinfo{title}{{Techniques of Model-Based Control}},
  \bibinfo{publisher}{Prentice Hall}, \bibinfo{year}{2002}.
\bibitem[{Majdpour et~al.(2021)Majdpour, Tsoukas, Yale, {El Fathi}, Rutkowski,
  Rene, Garfield, Legault, and Haidar}]{Majdpour2021}
\bibinfo{author}{D.~Majdpour}, \bibinfo{author}{M.~A. Tsoukas},
  \bibinfo{author}{J.~F. Yale}, \bibinfo{author}{A.~{El Fathi}},
  \bibinfo{author}{J.~Rutkowski}, \bibinfo{author}{J.~Rene},
  \bibinfo{author}{N.~Garfield}, \bibinfo{author}{L.~Legault},
  \bibinfo{author}{A.~Haidar},
\newblock \bibinfo{title}{{Fully Automated Artificial Pancreas for Adults With
  Type 1 Diabetes Using Multiple Hormones: Exploratory Experiments}},
\newblock \bibinfo{journal}{Canadian Journal of Diabetes} \bibinfo{volume}{45}
  (\bibinfo{year}{2021}) \bibinfo{pages}{734--742}.
  \DOIprefix\doi{10.1016/j.jcjd.2021.02.002}.
\bibitem[{Chase et~al.(2009)Chase, Lutz, Pencek, Zhang, and Porter}]{Chase2009}
\bibinfo{author}{H.~P. Chase}, \bibinfo{author}{K.~Lutz},
  \bibinfo{author}{R.~Pencek}, \bibinfo{author}{B.~Zhang},
  \bibinfo{author}{L.~Porter},
\newblock \bibinfo{title}{{Pramlintide Lowered Glucose Excursions and Was
  Well-Tolerated in Adolescents with Type 1 Diabetes: Results from a
  Randomized, Single-Blind, Placebo-Controlled, Crossover Study}},
\newblock \bibinfo{journal}{Journal of Pediatrics} \bibinfo{volume}{155}
  (\bibinfo{year}{2009}) \bibinfo{pages}{369--373}.
  \DOIprefix\doi{10.1016/j.jpeds.2009.03.012}.
\bibitem[{Colburn et~al.(1996)Colburn, Gottlieb, Koda, and
  Kolterman}]{Colburn1996}
\bibinfo{author}{W.~A. Colburn}, \bibinfo{author}{A.~B. Gottlieb},
  \bibinfo{author}{J.~Koda}, \bibinfo{author}{O.~G. Kolterman},
\newblock \bibinfo{title}{{Pharmacokinetics and Pharmacodynamics of AC137
  (25,28,29 Tripro-Amylin, Human) After Intravenous Bolus and Infusion Doses in
  Patients with Insulin-Dependent Diabetes}},
\newblock \bibinfo{journal}{The Journal of Clinical Pharmacology}
  \bibinfo{volume}{36} (\bibinfo{year}{1996}) \bibinfo{pages}{13--24}.
  \DOIprefix\doi{10.1002/j.1552-4604.1996.tb04147.x}.
\bibitem[{Kong et~al.(1998)Kong, Stubbs, King, Macdonald, Lambourne, Blackshaw,
  Perkins, and Tattersall}]{Kong1998}
\bibinfo{author}{M.~F. Kong}, \bibinfo{author}{T.~A. Stubbs},
  \bibinfo{author}{P.~King}, \bibinfo{author}{I.~A. Macdonald},
  \bibinfo{author}{J.~E. Lambourne}, \bibinfo{author}{P.~E. Blackshaw},
  \bibinfo{author}{A.~C. Perkins}, \bibinfo{author}{R.~B. Tattersall},
\newblock \bibinfo{title}{{The effect of single doses of pramlintide on gastric
  emptying of two meals in men with IDDM}},
\newblock \bibinfo{journal}{Diabetologia} \bibinfo{volume}{41}
  (\bibinfo{year}{1998}) \bibinfo{pages}{577--583}.
  \DOIprefix\doi{10.1007/s001250050949}.
\bibitem[{Heptulla et~al.(2005)Heptulla, Rodriguez, Bomgaars, and
  Raymond}]{Heptulla2005}
\bibinfo{author}{R.~A. Heptulla}, \bibinfo{author}{L.~M. Rodriguez},
  \bibinfo{author}{L.~Bomgaars}, \bibinfo{author}{M.~W. Raymond},
\newblock \bibinfo{title}{{The role of amylin and glucagon in the dampening of
  glycemic excursions in children with type 1 diabetes}},
\newblock \bibinfo{journal}{Diabetes} \bibinfo{volume}{54}
  (\bibinfo{year}{2005}) \bibinfo{pages}{1100--1107}.
  \DOIprefix\doi{10.2337/diabetes.54.4.1100}.
\bibitem[{Ahren et~al.(2002)Ahren, Adner, Svartberg, Petrella, Holst, and
  Gutniak}]{Ahren2002}
\bibinfo{author}{B.~Ahren}, \bibinfo{author}{N.~Adner},
  \bibinfo{author}{J.~Svartberg}, \bibinfo{author}{E.~Petrella},
  \bibinfo{author}{J.~J. Holst}, \bibinfo{author}{M.~K. Gutniak},
\newblock \bibinfo{title}{{Anti-diabetogenic effect of the human amylin
  analogue, pramlintide, in Type 1 diabetes is not mediated by GLP-1}},
\newblock \bibinfo{journal}{Diabetic Medicine} \bibinfo{volume}{19}
  (\bibinfo{year}{2002}) \bibinfo{pages}{790--792}.
  \DOIprefix\doi{10.1046/j.1464-5491.2002.00657\_1.x}.
\bibitem[{Edelman et~al.(2007)Edelman, Schroeder, and Frias}]{Edelman2007}
\bibinfo{author}{S.~V. Edelman}, \bibinfo{author}{B.~E. Schroeder},
  \bibinfo{author}{J.~P. Frias},
\newblock \bibinfo{title}{{Pramlintide acetate in the treatment of Type 2 and
  Type 1 diabetes mellitus}},
\newblock \bibinfo{journal}{Expert Review of Endocrinology \& Metabolism}
  \bibinfo{volume}{2} (\bibinfo{year}{2007}) \bibinfo{pages}{9--18}.
  \DOIprefix\doi{10.1586/17446651.2.1.9}.
\bibitem[{Kawamura et~al.(2015)Kawamura, Takamura, Hirose, Hashimoto,
  Higashide, Kashihara, Hashimura, and Shintaku}]{Kawamura2015}
\bibinfo{author}{T.~Kawamura}, \bibinfo{author}{C.~Takamura},
  \bibinfo{author}{M.~Hirose}, \bibinfo{author}{T.~Hashimoto},
  \bibinfo{author}{T.~Higashide}, \bibinfo{author}{Y.~Kashihara},
  \bibinfo{author}{K.~Hashimura}, \bibinfo{author}{H.~Shintaku},
\newblock \bibinfo{title}{{The factors affecting on estimation of carbohydrate
  content of meals in carbohydrate counting}},
\newblock \bibinfo{journal}{Clinical Pediatric Endocrinology}
  \bibinfo{volume}{24} (\bibinfo{year}{2015}) \bibinfo{pages}{153--165}.
  \DOIprefix\doi{10.1297/cpe.24.153}.
\bibitem[{Camerlingo et~al.(2021)Camerlingo, Vettoretti, {Del Favero},
  Facchinetti, and Sparacino}]{Camerlingo2021}
\bibinfo{author}{N.~Camerlingo}, \bibinfo{author}{M.~Vettoretti},
  \bibinfo{author}{S.~{Del Favero}}, \bibinfo{author}{A.~Facchinetti},
  \bibinfo{author}{G.~Sparacino},
\newblock \bibinfo{title}{{Mathematical Models of Meal Amount and Timing
  Variability With Implementation in the Type-1 Diabetes Patient Decision
  Simulator}},
\newblock \bibinfo{journal}{Journal of Diabetes Science and Technology}
  \bibinfo{volume}{15} (\bibinfo{year}{2021}) \bibinfo{pages}{346--359}.
  \DOIprefix\doi{10.1177/1932296820952123}.
\bibitem[{Battelino et~al.(2019)Battelino, Danne, Bergenstal, Amiel, Beck,
  Biester, Bosi, Buckingham, Cefalu, Close, Cobelli, Dassau, DeVries, Donaghue,
  Dovc, Doyle, Garg, Grunberger, Heller, Heinemann, Hirsch, Hovorka, Jia,
  Kordonouri, Kovatchev, Kowalski, Laffel, Levine, Mayorov, Mathieu, Murphy,
  Nimri, N{\o}rgaard, Parkin, Renard, Rodbard, Saboo, Schatz, Stoner, Urakami,
  Weinzimer, and Phillip}]{Battelino2019}
\bibinfo{author}{T.~Battelino}, \bibinfo{author}{T.~Danne},
  \bibinfo{author}{R.~M. Bergenstal}, \bibinfo{author}{S.~A. Amiel},
  \bibinfo{author}{R.~Beck}, \bibinfo{author}{T.~Biester},
  \bibinfo{author}{E.~Bosi}, \bibinfo{author}{B.~A. Buckingham},
  \bibinfo{author}{W.~T. Cefalu}, \bibinfo{author}{K.~L. Close},
  \bibinfo{author}{C.~Cobelli}, \bibinfo{author}{E.~Dassau},
  \bibinfo{author}{J.~H. DeVries}, \bibinfo{author}{K.~C. Donaghue},
  \bibinfo{author}{K.~Dovc}, \bibinfo{author}{F.~J. Doyle},
  \bibinfo{author}{S.~Garg}, \bibinfo{author}{G.~Grunberger},
  \bibinfo{author}{S.~Heller}, \bibinfo{author}{L.~Heinemann},
  \bibinfo{author}{I.~B. Hirsch}, \bibinfo{author}{R.~Hovorka},
  \bibinfo{author}{W.~Jia}, \bibinfo{author}{O.~Kordonouri},
  \bibinfo{author}{B.~Kovatchev}, \bibinfo{author}{A.~Kowalski},
  \bibinfo{author}{L.~Laffel}, \bibinfo{author}{B.~Levine},
  \bibinfo{author}{A.~Mayorov}, \bibinfo{author}{C.~Mathieu},
  \bibinfo{author}{H.~R. Murphy}, \bibinfo{author}{R.~Nimri},
  \bibinfo{author}{K.~N{\o}rgaard}, \bibinfo{author}{C.~G. Parkin},
  \bibinfo{author}{E.~Renard}, \bibinfo{author}{D.~Rodbard},
  \bibinfo{author}{B.~Saboo}, \bibinfo{author}{D.~Schatz},
  \bibinfo{author}{K.~Stoner}, \bibinfo{author}{T.~Urakami},
  \bibinfo{author}{S.~A. Weinzimer}, \bibinfo{author}{M.~Phillip},
\newblock \bibinfo{title}{{Clinical Targets for Continuous Glucose Monitoring
  Data Interpretation: Recommendations From the International Consensus on Time
  in Range}},
\newblock \bibinfo{journal}{Diabetes Care} \bibinfo{volume}{42}
  (\bibinfo{year}{2019}) \bibinfo{pages}{1593--1603}.
  \DOIprefix\doi{10.2337/dci19-0028}.
\bibitem[{Kovatchev(2019)}]{Kovatchev2019}
\bibinfo{author}{B.~Kovatchev},
\newblock \bibinfo{title}{{Glycemic Variability: Risk Factors, Assessment, and
  Control}},
\newblock \bibinfo{journal}{Journal of Diabetes Science and Technology}
  \bibinfo{volume}{13} (\bibinfo{year}{2019}) \bibinfo{pages}{627--635}.
  \DOIprefix\doi{10.1177/1932296819826111}.
\bibitem[{Bates et~al.(2015)Bates, M{\"{a}}chler, Bolker, and
  Walker}]{Bates2015}
\bibinfo{author}{D.~Bates}, \bibinfo{author}{M.~M{\"{a}}chler},
  \bibinfo{author}{B.~Bolker}, \bibinfo{author}{S.~Walker},
\newblock \bibinfo{title}{{Fitting Linear Mixed-Effects Models Using lme4}},
\newblock \bibinfo{journal}{Journal of Statistical Software}
  \bibinfo{volume}{67} (\bibinfo{year}{2015}).
  \DOIprefix\doi{10.18637/jss.v067.i01}.
\bibitem[{{R Core Team}(2021)}]{R2021}
\bibinfo{author}{{R Core Team}}, \bibinfo{title}{R: A Language and Environment
  for Statistical Computing}, \bibinfo{organization}{R Foundation for
  Statistical Computing}, \bibinfo{address}{Vienna, Austria},
  \bibinfo{year}{2021}. \URLprefix \url{https://www.R-project.org/}.
\bibitem[{Arel-Bundock et~al.(ming)Arel-Bundock, Greifer, and Heiss}]{ArelXX}
\bibinfo{author}{V.~Arel-Bundock}, \bibinfo{author}{N.~Greifer},
  \bibinfo{author}{A.~Heiss},
\newblock \bibinfo{title}{How to interpret statistical models using
  {marginaleffects} in {R} and {Python}},
\newblock \bibinfo{journal}{Journal of Statistical Software}
  (\bibinfo{year}{Forthcoming}).
\bibitem[{Huffman et~al.(2009)Huffman, McLean, and Seagrove}]{Huffman2009}
\bibinfo{author}{D.~M. Huffman}, \bibinfo{author}{G.~W. McLean},
  \bibinfo{author}{M.~A. Seagrove},
\newblock \bibinfo{title}{{Continuous subcutaneous pramlintide infusion therapy
  in patients with type 1 diabetes: observations from a pilot study.}},
\newblock \bibinfo{journal}{Endocrine practice : official journal of the
  American College of Endocrinology and the American Association of Clinical
  Endocrinologists} \bibinfo{volume}{15} (\bibinfo{year}{2009})
  \bibinfo{pages}{689--695}. \DOIprefix\doi{10.4158/EP09044.ORR1}.
\bibitem[{Andersen et~al.(2023)Andersen, Eloy, Famulla, Heise, Meiffren,
  Seroussi, Gaudier, Mégret, Chan, Soula, and Riddle}]{Andersen2023}
\bibinfo{author}{G.~Andersen}, \bibinfo{author}{R.~Eloy},
  \bibinfo{author}{S.~Famulla}, \bibinfo{author}{T.~Heise},
  \bibinfo{author}{G.~Meiffren}, \bibinfo{author}{C.~Seroussi},
  \bibinfo{author}{M.~Gaudier}, \bibinfo{author}{C.~Mégret},
  \bibinfo{author}{Y.~P. Chan}, \bibinfo{author}{O.~Soula},
  \bibinfo{author}{M.~Riddle},
\newblock \bibinfo{title}{A co-formulation of pramlintide and insulin a21g
  (ado09) improves postprandial glucose and short-term control of mean glucose,
  time in range, and body weight versus insulin aspart in adults with type 1
  diabetes},
\newblock \bibinfo{journal}{Diabetes, Obesity and Metabolism}
  \bibinfo{volume}{25} (\bibinfo{year}{2023}) \bibinfo{pages}{1241--1248}.
  \DOIprefix\doi{10.1111/dom.14972}.

\end{thebibliography}


\title{\\ \emph{Supplementary material for the manuscript}}

\author{Borja Pons Torres, Iván Sala-Mira, Clara Furió-Novejarque, Ricardo Sanz, Pedro García, José-Luis Díez, Jorge Bondia} 


\maketitle


\newpage

\onecolumn

\renewcommand{\thesection}{S\arabic{section}} \renewcommand{\thetable}{S\arabic{table}}

\setcounter{section}{0}
\setcounter{table}{0}

\section{Description of controller parameters}

This section includes Table \ref{tb:contr_par} with a description of all the relevant parameters used in the insulin-pramlintide control strategies in the manuscript.

\begin{table}[h]
  \caption{Control-related parameters and variables.}\label{tb:contr_par}
  \center
  \resizebox{0.91\textwidth}{!}{
  \begin{tabular}{ c  c  l  }  
      \toprule
      Symbol & Units & Description \\
      \midrule
      $s$ & - & Laplace variable \\
      $k$ & - & Discrete sampling instant \\
      $\hat{d}(s)$ & g & Estimated meal disturbance \\
      $u^I(s)$ & U & Incremental insulin infusion rate (deviation from the basal insulin infusion) \\
      $G(s)$ & mg/dL & Current blood glucose measured by a continuous glucose monitoring device \\
      $G_b$ & mg/dL & Basal glucose value \\
      $y(s)$ & mg/dL & Incremental blood glucose value with respect to the basal glucose value \\
      $\gamma$ & U/g & Positive constant defining the controller gain \\
      $F(s)$ & - & Filter used to estimate the disturbance \\
      $\theta$ & - & Time constant of $F(s)$ \\
      $\tilde{G}_i(s)$ & - & Nominal transfer function describing the glycemic effect of $i$ \\
                       &   & (with $i$ being either a meal, $d$, or insulin, $u$) \\
      $\tau_{1i}$ & min & First time constant for the transfer function describing the glycemic effect of $i$\\
      $\tau_{2i}$ & min & Second time constant for the transfer function describing the glycemic effect of $i$\\
      $K_i^j$ & - & Positive gain of the effect of $i$, tailored to the patient $j$ \\
      $u^B$ & U & Prandial insulin bolus \\
      $\widehat{CHO}$ & g & Meal carbohydrate content estimated by the patient \\
      CIR & g/U     & Carbohydrate to insulin ratio \\
      $\nu$ & - & Attenuating safety factor (set to 0.8) \\
      $p(k)$ & {\microg} & Total pramlintide value delivered in the $k$-th iteration \\
      $p^B(k)$ & {\microg} & Pramlintide delivered as boluses in the $k$-th iteration \\
      $p^I(k)$ & {\microg}/U & Pramlintide delivered as continuous infusion in the $k$-th iteration \\
      $u(k)$ &   U & Total insulin value delivered in the $k$-th iteration \\
      $u^B(k)$ & U & Insulin delivered as boluses in the $k$-th iteration \\
      $u^I(k)$ & U & Insulin delivered as continuous infusion in the $k$-th iteration \\  
      $u_b(k)$ & U & Basal insulin value in the $k$-th iteration \\  
      $m_u(k)$ & U/s & Insulin slope in the $k$-th iteration \\  
      $m_g(k)$ & mg/dL$\cdot$s & Glucose slope in the $k$-th iteration \\  
      $\lambda$ & {\microg} & Pramlintide bolus value (pramlintide dosing Strategy 1) \\
      $B_1(k)$ & - & First condition for pramlintide delivery in the $k$-th iteration (Strategy 1)\\
      $B_2(k)$ & - & Second condition for pramlintide delivery in the $k$-th iteration (Strategy 1)\\
      $z_1$   & -  & Constant defining the insulin delivery threshold in condition $B_2(k)$ (Strategy 1)\\ 
      $z_2$   & -  & Constant defining the glucose slope threshold in condition $B_2(k)$ (Strategy 1)\\
      $z_3$   & -  & Constant defining the iteration threshold in condition $B_2(k)$ (Strategy 1)\\
      $\Delta  k_p$ & - & Number of iterations between the current and the last administration of pramlintide\\
      $\rho$ & {\microg}/U & Pramlintide infusion delivery ratio (Strategy 2)\\
      $\varphi $ & {\microg} & Pramlintide bolus value (Strategy 3)\\
      $\delta$ & {\microg}/U & Pramlintide bolus and infusion delivery ratio (Strategy 4)\\ 
        \bottomrule
  \end{tabular}
  }
\end{table}

\newpage

\section{In silico validation scenario}

The 14-day scenario was generated by drawing meals from the distributions of mealtime and carbohydrate amounts described in \citep{Camerlingo2021}, representing real-life meal patterns. Up to seven meals can be included in a day: three main meals (breakfast, lunch, dinner) and a varying number of snacks (up to four), resulting in 168.58 (57.14)~g daily carbohydrates (median (standard deviation)). The resulting meals carbohydrate content and time values are listed in Table \ref{tb:scen}.

\begin{table}[h]
  \caption{Meal and snacks inputs for the validation scenario.}\label{tb:scen}
  \center
  \resizebox{0.88\textwidth}{!}{
  \begin{tabular}{ c  c  c  c |  c  c  c  c }  
      \toprule
      Day & Type of meal & Time & Meal amount & Day & Type of meal & Time & Meal amount \\
      \midrule
      \multirow{7}{*}{1} & Breakfast & 10:05 h &  64.88 g & \multirow{3}{*}{8} & Breakfast & 10:30 h &  60.55 g \\ 
      &     Snack & 11:50 h &  47.69 g                    & &     Lunch & 15:45 h &  79.69 g \\ 
      &     Snack & 12:10 h &  40.61 g                    &  &    Dinner & 22:10 h &  36.17 g \\ 
      &     Snack & 14:15 h &  13.40 g                    &  & & & \\
      &     Lunch & 14:55 h &  54.05 g                    & \multirow{6}{*}{9} &     Snack & 00:30 h &  20.00 g \\ 
      &     Snack & 18:40 h &  52.74 g                    & &     Snack & 02:15 h &  65.22 g \\ 
      &    Dinner & 22:20 h &  33.95 g                    &  & Breakfast & 09:45 h &  39.51 g \\ 
     & & &                                                & &     Lunch & 16:00 h &  21.36 g \\
     \multirow{4}{*}{2} & Breakfast & 07:25 h &  41.87 g  & &     Snack & 17:50 h &   4.27 g \\  
      &     Lunch & 09:55 h &  84.38 g                    & &     Snack & 20:30 h &  50.06 g \\ 
      &     Snack & 12:45 h &  13.13 g                    & & & & \\ 
      &    Dinner & 15:05 h &  24.14 g                    & \multirow{4}{*}{10} &    Dinner & 00:30 h &  77.96 g \\ 
     & & &                                                & & Breakfast & 07:20 h &  60.08 g \\ 
     \multirow{3}{*}{3} & Breakfast & 06:00 h &  55.08 g  &  &     Lunch & 11:25 h &  32.72 g \\  
      &     Lunch & 12:30 h &  61.44 g                    & &    Dinner & 18:40 h &  54.45 g \\ 
      &    Dinner & 20:25 h &  41.17 g                    &  & & & \\ 
     & & &                                                & \multirow{3}{*}{11} & Breakfast & 10:10 h &  33.18 g \\
     \multirow{3}{*}{4} & Breakfast & 09:30 h &  30.89 g  & &     Lunch & 12:55 h &  23.17 g \\  
      &     Lunch & 14:00 h & 102.01 g                    &  &    Dinner & 20:25 h &  28.11 g \\ 
      &    Dinner & 20:35 h &  41.44 g                    &  & & & \\ 
     & & &                                                & \multirow{4}{*}{12} & Breakfast & 09:20 h &  15.29 g \\ 
     \multirow{5}{*}{5} & Breakfast & 06:45 h &  16.65 g  & &     Lunch & 15:00 h &  34.60 g \\  
      &     Snack & 08:55 h &  30.40 g                    & &     Snack & 19:10 h &   8.03 g \\ 
      &     Lunch & 13:20 h &  38.77 g                    & &    Dinner & 22:35 h &  53.85 g \\  
      &    Dinner & 20:05 h &  23.73 g                    &  & & & \\ 
      &     Snack & 21:30 h &   6.23 g                    & \multirow{5}{*}{13} & Breakfast & 06:40 h &  50.85 g \\  
     & & &                                                & &    Snack & 08:15 h &  22.25 g \\ 
     \multirow{5}{*}{6} & Breakfast & 07:25 h &  33.45 g  & &    Lunch & 10:50 h &  28.39 g \\  
      &     Lunch & 12:25 h &  22.03 g                    & &     Snack & 14:15 h &  42.92 g \\ 
      &    Dinner & 19:15 h &  73.75 g                    & &    Dinner & 18:45 h &  75.73 g \\ 
      &     Snack & 21:40 h &   3.67 g                    & & & & \\ 
      &     Snack & 22:25 h &   6.36 g                    & \multirow{3}{*}{14} & Breakfast & 09:45 h &  47.73 g \\ 
     & & &                                                &  &     Lunch & 15:10 h &  43.66 g \\
     \multirow{5}{*}{7} &     Snack & 01:15 h &   4.12 g  & &    Dinner & 21:20 h &  69.27 g \\ 
      & Breakfast & 06:30 h &  22.97 g  & & & \\ 
      &     Snack & 07:50 h &  17.73 g  & & & \\ 
      &     Lunch & 10:10 h &  30.74 g  & & & \\ 
      &    Dinner & 15:55 h &  47.58 g  & & & \\ 
      \bottomrule
  \end{tabular}
  }
\end{table}


\end{document}